\documentclass[12pt]{article}
\pdfoutput=1

\usepackage[T1]{fontenc}

\usepackage{amsmath}
\usepackage{amssymb}
\usepackage{slashed}
\usepackage{tikz}
\usepackage{ulem}
\usepackage{nameref}
\usepackage[numbers,sort&compress]{natbib}
\usepackage{hyperref}
\usepackage{cleveref}
\crefname{equation}{Eq.}{Eqs.}
\crefname{figure}{Fig.}{Figs.}
\crefname{table}{Table}{Tables}
\crefname{section}{Section}{Sections}

\newcommand\scalemath[2]{\scalebox{#1} {\mbox{\ensuremath{\displaystyle #2}}}}
\usepackage[letterpaper,margin=1in,bottom=1in]{geometry}
\usepackage{float} % allows forcing plot and table locations with [H]
\usepackage{parskip} % skip spaces between paragraphs
\usepackage{tabulary} % table environment with auto word wrapping
\usepackage{soul} % strikeout for tracking edits
\usepackage{subfigure}
\usepackage{graphicx}
\usepackage[section]{placeins} % keep figures inside their sections

\def\lhc2{LHC~Run~II}

\newcommand{\code}[1]{\texttt{#1}}

\newcommand{\ifb}{~\textrm{fb}^{-1}}

\def\beq{\begin{equation}}
\def\be{\begin{equation}}
\def\beqn{\begin{eqnarray}}
\def\ee{\end{equation}}
\def\eeq{\end{equation}}
\def\eeqn{\end{eqnarray}}

\def\checkmark{\tikz\fill[scale=0.4](0,.35) -- (.25,0) -- (1,.7) -- (.25,.15) -- cycle;}
\begin{document}

\author{Amin Aboubrahim$^b$\footnote{Email: a.abouibrahim@northeastern.edu}~, Wan-Zhe Feng$^a$\footnote{Email: vicf@tju.edu.cn}~~and Pran Nath$^b$\footnote{Email: p.nath@northeastern.edu}\\~\\
$^{a}$ \textit{\normalsize Center for Joint Quantum Studies and Department of Physics,}\\
\textit{\normalsize School of Science, Tianjin University, Tianjin 300350, PR. China}\\
$^{b}$ \textit{\normalsize Department of Physics, Northeastern University,
Boston, MA 02115-5000, USA} \\}

\title{\vspace{-2.0cm}\begin{flushright}
{\small CJQS-2019-027}
\end{flushright}
{A long-lived stop with freeze-in and freeze-out dark matter in the hidden sector}}

\date{}
\maketitle

\begin{abstract}
In extended supersymmetric models with a hidden sector the lightest $R$-parity odd particle
can reside in the hidden sector and act as dark matter.
We consider the case when the hidden sector has ultraweak interactions with the visible sector.
An interesting phenomenon arises if the LSP of the visible sector
is charged in which case it will decay to the hidden sector dark matter.
Due to the ultraweak interactions, the LSP of the visible sector will be
long-lived decaying outside the detector after leaving a track inside.
We investigate this possibility in the framework of a
$U(1)_X$-extended MSSM/SUGRA model with a small gauge kinetic mixing and mass mixing between the $U(1)_X$ and $U(1)_Y$ where $U(1)_Y$ is the gauge group of the hypercharge.
Specifically we investigate the case when the LSP of MSSM is a stop
which decays into the hidden sector dark matter and
has a lifetime long enough to traverse the LHC detector without decay.
It is shown that such a particle can be detected at the HL-LHC and HE-LHC
as an $R$-hadron which will look like a slow moving muon
with a large transverse momentum $p_T$ and so can be detected
by the track it leaves in the inner tracker and in the muon spectrometer.
Further, due to the ultraweak couplings between the hidden sector and the MSSM fields,
the dark matter particle has a relic density arising from a combination of the
freeze-out and freeze-in mechanisms.
It is found that even for the ultraweak or feeble interactions the freeze-out contribution
relative to freeze-in contribution to the relic density is substantial to dominant,
varying between 30\% to 74\% for the model points considered.
It is subdominant to freeze-in for relatively small stop masses
with relatively larger stop annihilation cross-sections and
the dominant contribution to the relic density for relatively large stop masses
and relatively smaller stop annihilation cross-sections.
Our analysis shows that the freeze-out contribution must be included for any
realistic analysis even for dark matter particles
with ultraweak or feeble interactions with the visible sector.
A discovery of a long-lived stop as the lightest particle of the MSSM
may point to the nature of dark matter and its production mechanism in the early universe.
\end{abstract}

\section{Introduction}\label{sec:intro}

The experiment at the Large Hadron Collider (LHC) has so far analyzed up to 139$\ifb$ of data
for each of ATLAS and CMS and the results are consistent with the Standard Model (SM).
Specifically there is yet no signal for supersymmetry.
The lack of observation of supersymmetry (SUSY) is not surprising in view of the measurement of
the Higgs boson mass at 125 GeV~\cite{Aad:2012tfa,Chatrchyan:2012ufa}
which indicates that the size of weak scale supersymmetry lies in the TeV region.
Thus the SUSY parameter space giving rise to traditional signals which involve final states
with large missing energy due to a neutralino as the lightest supersymmetric particle (LSP),
or hard jets arising from the decay of strongly interacting SUSY particles (squarks and gluinos)
or high momentum leptons coming from the decay of electroweak gauginos
is now signficantly more constrained.
Constraints are less severe for more rare processes because of their small production cross-sections.
Even for the largest production cross-sections the region of compressed sparticle spectrum
is as yet not significantly constrained.
In the $R$-parity conserving minimal supersymmetric standard model (MSSM)
the decay chain always ends up with the LSP along with standard model particles.
If the mass gap between the produced sparticle and the LSP is small
(in which case the sparticle is the next-to-LSP, or NLSP),
the decay products are soft and thus pose a challenge to experiment at the LHC.
This region also requires attention regarding satisfaction of relic density constraints.
For instance, in the region where the stau mass is close to the mass of the LSP
the relic density is controlled by coannihilation between the stau and the neutralino
(for a recent analysis see~\cite{Aboubrahim:2017aen} and references therein).
The coannihilation region is particularly useful for models where the LSP is bino-like
and the relic density arising from LSP annihilation alone would be
far in excess of the observed relic density.
Here one needs coannihilation to deplete the LSP relic density to
the experimentally observed value.
Aside from the stau, a gluino, a stop, or a chargino can be the particles
that coannihilate with the LSP (for recent works on gluino, stop and chargino as
coannihilating particles see Refs.~\cite{Feldman:2009zc,Nath:2016kfp,Kaufman:2015nda,Aboubrahim:2017wjl} and the references therein).

Another search which is  still not highly constrained is that for exotic signals, in particular, long-lived particles. Most long-lived particle searches at the LHC consider an NLSP very close in mass to the LSP ($\Delta m\sim$ few GeV down to MeV) resulting in a highly suppressed phase space. This leads to a small decay width for the NLSP and thus a long-lived particle. If the particle is charged and stable over detector length it can be identified by the track it leaves in the inner tracker and in the muon spectrometer. Other signatures are possible such as a disappearing track where a charged particle can decay into very soft final states which escape the trigger threshold (for a good review of collider searches for long-lived particles, see Refs.~\cite{Lee:2018pag,Alimena:2019zri}). Thus, ATLAS and CMS were not designed to look for long-lived particles and part of the upcoming upgrade is to further the capabilities of these detectors to become more sensitive to such searches.

Long-lived particles can arise in SUSY models with a hidden sector if the hidden sector has ultraweak interactions with the visible sector and the LSP of the visible sector decays into the hidden sector.
In this work we discuss an MSSM/SUGRA (supergravity) model extended by an extra $U(1)_X$ gauge group with a gauge kinetic mixing~\cite{Holdom:1985ag,Holdom:1991} and Stueckelberg mass mixing~\cite{Kors:Nath,st-mass-mixing,WZFPN,Feldman:2007wj} between the $U(1)_X$ and the SM hypercharge $U(1)_Y$ gauge groups. The model contains additional chiral scalar
superfields $S$  and $\bar S$ and a vector superfield $C$. The fermionic component of $S$ and $\bar S$ and the gaugino components of $C$ mix with the MSSM neutralino fields producing a $6\times 6$ neutralino mass matrix. The input mass hierarchy of the neutralino sector allows us to have the LSP as the neutralino of the hidden sector. Thus, the decay of the NLSP or any other MSSM field into the hidden sector LSP is highly suppressed by our choice of the gauge kinetic and mass mixing parameters. Being a dark matter candidate and possessing very weak interactions with the visible sector, the LSP will be produced out of equilibrium in the early universe. For
MSSM coupled to the hidden sector by ultraweak interactions, the LSP relic density cannot be accounted for by the usual freeze-out mechanism alone.
However, it is shown that the dark matter relic density consistent with
experiment can be achieved by a combination of the freeze-out\footnote{As will be explained in more details in Section~\ref{sec:DM},
the freeze-out contribution is not from the leftover of dark matter after the annihilation into visible sector particles, but arises from the decay of the NLSP (in our case, stop) after it freezes out.}
and freeze-in~\cite{Hall:2009bx,Belanger:2018mqt,Tsao:2017vtn} mechanisms.
Despite the small decay widths of all heavier visible sector sparticles into the LSP, this decay will eventually happen over a period of time thus producing the desired
contribution to the relic abundance.

In the analysis here
we consider a set of benchmarks satisfying the constraints on the Higgs boson mass and the relic density as measured by the Planck Collaboration~\cite{Aghanim:2018eyx} where the stop is the NLSP and a long-lived particle. We perform a collider analysis discussing the prospects of discovering a long-lived stop at HL-LHC and HE-LHC~\cite{CidVidal:2018eel,Cepeda:2019klc,Benedikt:2018ofy,Zimmermann:2018koi} (for previous works on HL-LHC and HE-LHC, see Refs.~\cite{Aboubrahim:2018bil,Aboubrahim:2018tpf,Aboubrahim:2019qpc,Aboubrahim:2019vjl,Aboubrahim:2019mxn}). The stop has very late decays into the hidden sector LSP but is stable over detector length and so can be identified by the track it leaves in the detector after hadronizing into what is known as an $R$-hadron.  We note that several works exist in the literature on supersymmetric $U(1)$ extensions of MSSM and
their implications on dark matter and collider searches (see, e.g.,~\cite{U1extensions}). Also, several works on signatures of long-lived particles at colliders with freeze-in dark matter have appeared recently~\cite{Co:2015pka,Chakraborti:2019ohe,Belanger:2018sti,No:2019gvl,Calibbi:2018fqf,Banerjee:2018uut} as well as scenarios testing for freeze-in via direct detection~\cite{Heeba:2019jho,Hambye:2018dpi,Mohapatra:2019ysk,Bernal:2018ins,Bernal:2018kcw} and indirect detection~\cite{Bernal:2017kxu,Heikinheimo:2018duk}.  The analysis of this work is significantly different from these.

The outline of the rest of the  paper is as follows: in Section~\ref{sec:model} we give an overview of the $U(1)_X$-extended MSSM/SUGRA model used in this work
followed by a discussion of freeze-in dark matter relic density  in Section~\ref{sec:DM}.  A discussion of the high scale model
 input and benchmarks  is given in Section~\ref{sec:implement} and production of stops at the LHC along with their cross-sections given
 in Section~\ref{sec:production}. Signal and SM background simulation along with the adopted selection criteria and results are discussed in Sections~\ref{sec:simulation} and~\ref{sec:results}. Conclusions are given in Section~\ref{sec:conc}.

\section{The model}\label{sec:model}

As discussed above we consider an extension of the standard model gauge group by
an additional abelian gauge group $U(1)_X$. The particle spectrum  in the visible sector, i.e., quarks, leptons, Higgs and their superpartners are assumed neutral under $U(1)_X$. We focus first on the abelian gauge sector of the extended model which contains two $U(1)$ vector superfields, i.e.,
a vector superfield $B$ associated with the hypercharge gauge group $U(1)_Y$, a vector superfield $C$ associated with the hidden sector gauge group $U(1)_X$.
In the Wess-Zumino gauge the $B$ and $C$ superfields have the following components
\begin{equation}
B=-\theta\sigma^{\mu}\bar{\theta}B_{\mu}+i\theta\theta\bar{\theta}\bar{\lambda}_B-i\bar{\theta}\bar{\theta}\theta\lambda_B+\frac{1}{2}\theta\theta\bar{\theta}\bar{\theta}D_B,
\end{equation}
and
\begin{align}
C=-\theta\sigma^{\mu}\bar{\theta}C_{\mu}+i\theta\theta\bar{\theta}\bar{\lambda}_{C}-i\bar{\theta}\bar{\theta}\theta\lambda_{C}+\frac{1}{2}\theta\theta\bar{\theta}\bar{\theta}D_{C}.
\end{align}
The gauge kinetic energy  sector of the model is
\begin{equation}
\mathcal{L}_{\rm gk}=-\frac{1}{4}(B_{\mu\nu}B^{\mu\nu}+C_{\mu\nu}C^{\mu\nu})-i\lambda_B\sigma^{\mu}\partial_{\mu}\bar{\lambda}_B-i\lambda_{C}\sigma^{\mu}\partial_{\mu}\bar{\lambda}_{C}+\frac{1}{2}(D^2_B+D^2_{C}).
\label{kinetic-1}
\end{equation}
Next we allow  gauge kinetic mixing between the $U(1)_X$ and $U(1)_Y$ sectors through terms of the form
\begin{equation}
-\frac{\delta}{2}B^{\mu\nu}C_{\mu\nu}-i\delta(\lambda_{C}\sigma^{\mu}\partial_{\mu}\bar{\lambda}_B+\lambda_{B}\sigma^{\mu}\partial_{\mu}\bar{\lambda}_{C})+\delta D_B D_{C}.
\label{kinetic-2}
\end{equation}
As a result of Eq.~(\ref{kinetic-2}) the hidden $U(1)_X$ interacts with the MSSM fields via the kinetic mixing parameter $\delta$
which can be chosen to be very small. The kinetic terms in Eq.~(\ref{kinetic-1}) and Eq.~(\ref{kinetic-2})  can be diagonalized using the transformation
\beqn
\left(\begin{matrix} B^{\mu} \cr
C^{\mu}
\end{matrix}\right) = \left(\begin{matrix} 1 & -s_{\delta} \cr
0 & c_{\delta}
\end{matrix}\right)\left(\begin{matrix} B'^{\mu} \cr
C'^{\mu}
\end{matrix}\right),
\label{rotation}
\eeqn
where $c_{\delta}=1/(1-\delta^2)^{1/2}$ and $s_{\delta}=\delta/(1-\delta^2)^{1/2}$.

Aside from gauge kinetic mixing, we assume a Stueckelberg mass mixing between the $U(1)_X$ and $U(1)_Y$ sectors so that
\begin{equation}
\mathcal{L}_{\rm St}=\int d\theta^2 d\bar{\theta}^2(M_1 C+M_2 B+S+\bar{S})^2,
\label{lag}
\end{equation}
where $S$ and $\bar S$ are chiral superfields.
$M_1$ is the mass of the hidden sector field $C$ when $M_2=0$,
and $M_2$ gives the mixing between hidden sector field and the hypercharge field $B$.
We note that Eq.~(\ref{lag}) is invariant under $U(1)_Y$ and $U(1)_X$ gauge transformations so that,
\begin{align}
&\delta_Y B = \Lambda_Y+\bar{\Lambda}_Y, \, \, \, \, \, \delta_Y S = -M_2\Lambda_Y, \\ \nonumber
& \delta_X C = \Lambda_X+\bar{\Lambda}_X, \, \, \, \delta_X S = -M_1\Lambda_X,
\end{align}
with $\delta_X B=0$ and $\delta_Y C=0$ implying the invariance of $B$ and $C$ under $U(1)_X$ and $U(1)_Y$, respectively.
The chiral scalar superfield  $S$ has the expansion in component form so that
\begin{align}
S = &\frac{1}{2}(\rho+i a)+\theta\chi+i\theta\sigma^{\mu}\bar{\theta}\frac{1}{2}(\partial_{\mu}\rho+i\partial_{\mu}a) \\ \nonumber
&+\theta\theta F+\frac{i}{2}\theta\theta\bar{\theta}\bar{\sigma}^{\mu}\partial_{\mu}\chi+\frac{1}{8}\theta\theta\bar{\theta}\bar{\theta}(\square\rho+i\square a),
\end{align}
and a similar expansion holds for $\bar S$. Further, in component notation, $\mathcal{L}_{\rm St}$ is given by
\begin{align}
\mathcal{L}_{\rm St} = &-\frac{1}{2}(M_1 C_{\mu}+M_2 B_{\mu}+\partial_{\mu}a)^2-\frac{1}{2}(\partial_{\mu}\rho)^2-i\chi\sigma^{\mu}\partial_{\mu}\bar{\chi}+2|F|^2 \\ \nonumber
&+\rho(M_1 D_{C}+M_2 D_B)+\bar{\chi}(M_1\bar{\lambda}_{C}+M_2\bar{\lambda}_B)+\chi(M_1\lambda_{C}+M_2\lambda_B).
\end{align}
In the unitary gauge the axion field $a$ is absorbed to generate mass for the $U(1)_X$ gauge boson.
\\\\
It is convenient from this point on to introduce Majorana spinors $\psi_S$, $\lambda_X$ and $\lambda_Y$ so that
 \begin{equation}
  \psi_S =
  \begin{pmatrix}
    \chi_{\alpha}  \\
    \bar{\chi}^{\dot{\alpha}}
  \end{pmatrix},\quad
  \lambda_X=
  \begin{pmatrix}
    \lambda_{C\alpha}  \\
    \bar{\lambda}^{\dot{\alpha}}_{C}
  \end{pmatrix},\quad
  \lambda_Y=
  \begin{pmatrix}
    \lambda_{B\alpha}  \\
    \bar{\lambda}^{\dot{\alpha}}_{B}
  \end{pmatrix}.
  \label{spinors}
\end{equation}
In addition to the above we add soft terms to the Lagrangian so that
\begin{equation}
\Delta\mathcal{L}_{\rm soft} \
=-\left(\frac{1}{2}m_X\bar{\lambda}_X\lambda_X+ M_{XY}\bar{\lambda}_X\lambda_Y\right)-\frac{1}{2} m^2_{\rho}\rho^2,
\end{equation}
where $m_X$ is mass of the $U(1)_X$ gaugino and $M_{XY}$ is the $U(1)_X$-$U(1)_Y$ gaugino
mixing mass.
We note that the mixing parameter $M_{XY}$ and $M_2$ even when set to zero at the grand unification scale will assume
non-vanishing values due  to renormalization group evolution. Thus $M_{XY}$ has the beta-function evolution so that
\begin{equation}
\beta^{(1)}_{M_{XY}}=\frac{33}{5}g^2_Y\left[M_{XY}-(M_1+m_X)s_{\delta}+M_{XY}s^2_{\delta} \right],
\end{equation}
where $g_Y$ is the $U(1)_Y$ gauge coupling. Similarly, the  mixing parameter $M_2$
has the beta-function so that
\begin{equation}
\beta^{(1)}_{M_2}=\frac{33}{5}g^2_Y(M_2-M_1 s_{\delta}).
\label{m2rge}
\end{equation}
In the MSSM sector we will take the soft terms to consist of $m_0, ~A_0, ~m_1, ~m_2, ~m_3, ~\tan\beta,
 ~\text{sgn} (\mu)$. Here
 $m_0$ is the universal scalar mass, $A_0$ is the universal trilinear coupling, $m_1,  ~m_2,  ~m_3$ are the masses of the $U(1)_Y$, $SU(2)_L$, and $SU(3)_C$ gauginos, $\tan\beta=v_u/v_d$ is the ratio of the Higgs vacuum expectation values and $\text{sgn}(\mu)$ is the sign of the Higgs mixing parameter which is chosen to be positive. Here we have assumed non-universalities in the gaugino mass sector which will be useful in the analysis
 in Section \ref{sec:implement} (for some relevant works on non-universalities in the gaugino masses
 see Ref.~\cite{nonuni-gaugino}).

We focus first on the neutralino sector of the extended SUGRA model. We choose as basis $(\psi_S,\lambda_X,\lambda_Y,\lambda_3,\tilde h_1, \tilde h_2)$ where the first two fields arise from the extended sector and the last four, i.e.,
 $\lambda_Y, \lambda_3, \tilde h_1, \tilde h_2$ are the gaugino and higgsino fields of the MSSM sector. Using Eq.~(\ref{rotation}) we rotate into the new basis $(\psi_S,\lambda'_X,\lambda'_Y,\lambda_3,\tilde h_1, \tilde h_2)$ so that the $6\times 6$ neutralino mass matrix takes the form
\beqn
\scalemath{0.9}{
\left(\begin{matrix}  0 & M_1 c_{\delta}-M_2 s_{\delta} & M_2 & 0 & 0 & 0 \cr
M_1 c_{\delta}-M_2 s_{\delta} & m_X c^2_{\delta}+m_1 s^2_{\delta}-M_{XY}c_{\delta}s_{\delta} & -m_1 s_{\delta}+M_{XY}c_{\delta} & 0 & s_{\delta}c_{\beta}s_W M_Z & -s_{\delta}s_{\beta}s_W M_Z \cr
M_2 & -m_1 s_{\delta}+M_{XY}c_{\delta} & m_1 & 0 & -c_{\beta}s_W M_Z & s_{\beta}s_W M_Z \cr
0 & 0 & 0 & m_2 & c_{\beta}c_W M_Z & -s_{\beta}c_W M_Z \cr
0 & s_{\delta}c_{\beta}s_W M_Z & -c_{\beta}s_W M_Z & c_{\beta}c_W M_Z & 0 & -\mu \cr
0 & -s_{\delta}s_{\beta}s_W M_Z & s_{\beta}s_W M_Z & -s_{\beta}c_W M_Z & -\mu & 0 \cr
\end{matrix}\right)},
\eeqn
where
$s_{\beta}\equiv\sin\beta$, $c_{\beta}\equiv\cos\beta$, $s_W\equiv\sin\theta_W$, $c_W\equiv\cos\theta_W$ with $M_Z$ being the $Z$ boson mass and $\theta_W$ the Weinberg mixing angle.
 We label the mass eigenstates as
\begin{equation}
\tilde\xi^0_1, ~\tilde\xi^0_2; ~\tilde \chi_1^0, ~\tilde \chi_2^0, ~\tilde \chi_3^0, ~\tilde \chi_4^0\,.
\end{equation}
Since the mixing parameter $\delta$ is very small,
the first two neutralinos $\tilde\xi^0_1$ and $\tilde\xi^0_2$ reside mostly in the hidden sector while the remaining four $\tilde \chi_i^0$
($i=1\cdots 4$) reside mostly in the MSSM sector.
In the limit of small mixings between the hidden and the MSSM sectors the masses of the hidden sector neutralinos are
\begin{equation}
m_{\tilde\xi^0_1}=\sqrt{M_1^2+\frac{1}{4}\tilde m^2_X}-\frac{1}{2}\tilde m_X, \quad \text{and} \quad m_{\tilde\xi^0_2}=\sqrt{M_1^2+\frac{1}{4}\tilde m^2_X}+\frac{1}{2}\tilde m_X.
\end{equation}
For the case when the lighter hidden neutralino $\tilde \xi_1^0$ is the least massive of all sparticles in the $U(1)_X$-extended SUGRA model, $\tilde\xi^0_1$ is the LSP and thus the dark matter candidate.
Such a possibility has been foreseen in previous works (see, e.g.,~\cite{Feldman:2006wd,Feldman:2008xs,Feldman:2009wv}).

We turn now to the charge neutral gauge vector boson sector. Here the $2\times 2$ mass-squared matrix of the standard model
is enlarged to become a $3\times 3$ mass-squared matrix in the $U(1)_X$-extended SUGRA model.
Thus  after spontaneous electroweak symmetry breaking and  the Stueckelberg mass growth the
$3\times 3$ mass-squared matrix of neutral vector bosons in the basis $(C'_{\mu}, B'_{\mu}, A^3_{\mu})$ is given by
\beqn
\mathcal{M}^2_V=\left(\begin{matrix}  M_1^2\kappa^2+\frac{1}{4}g^2_Y v^2 s^2_{\delta} & M_1 M_2\kappa-\frac{1}{4}g^2_Y v^2 s_{\delta} & \frac{1}{4}g_Y g_2 v^2 s_{\delta} \cr
M_1 M_2\kappa-\frac{1}{4}g^2_Y v^2 s_{\delta} & M_2^2+\frac{1}{4}g^2_Y v^2 & -\frac{1}{4}g_Y g_2 v^2 \cr
\frac{1}{4}g_Y g_2 v^2 s_{\delta} & -\frac{1}{4}g_Y g_2 v^2 & \frac{1}{4}g^2_2 v^2 \cr
\end{matrix}\right),
\label{zmassmatrix}
\eeqn
where $A^3_{\mu}$ is the third isospin component, $g_2$ is the $SU(2)_L$ gauge coupling, $\kappa=(c_{\delta}-\epsilon s_{\delta})$, $\epsilon=M_2/M_1$ and $v^2=v^2_u+v^2_d$. The mass-squared matrix of Eq.~(\ref{zmassmatrix}) has one zero eigenvalue which is the photon while the other two eigenvalues are
\begin{align}
&M^2_{\pm} = \frac{1}{2}\Bigg[M_1^2\kappa^2+M^2_2+\frac{1}{4}v^2[g_Y^2 c^2_{\delta}+g_2^2] \nonumber \\
&\pm \sqrt{\left(M_1^2\kappa^2+M^2_2+\frac{1}{4}v^2[g_Y^2 c^2_{\delta}+g_2^2]\right)^2-\Big[M_1^2 g_2^2v^2\kappa^2+M_1^2g^2_Yv^2 c^2_{\delta}+M_2^2g^2_2 v^2\Big]}~\Bigg],
\label{bosons}
\end{align}
where $M_+$ is identified as the $Z'$ boson mass while $M_-$ as  the $Z$ boson. The diagonalization of the mass-squared matrix of Eq.~(\ref{zmassmatrix}) can be done via two orthogonal transformations where the first is given by~\cite{Feldman:2007wj}
\beqn
\mathcal{O}=\left(\begin{matrix} 1/c_{\delta} & -s_{\delta}/c_{\delta} & 0 \cr
s_{\delta}/c_{\delta} & 1/c_{\delta} & 0 \cr
0 & 0 & 1 \cr
\end{matrix}\right),
\label{omatrix}
\eeqn
which transforms the mass matrix to $\mathcal{M'}^2_V=\mathcal{O}^{T}\mathcal{M}^2_V\mathcal{O}$,
\beqn
\mathcal{M'}^2_V=\left(\begin{matrix}  M_1^2 & M_1^2\alpha & 0 \cr
M_1^2\alpha & M_1^2\alpha^2+\frac{1}{4}g^2_Y v^2 c^2_{\delta} & -\frac{1}{4}g_Y g_2 v^2 c_{\delta} \cr
0 & -\frac{1}{4}g_Y g_2 v^2 c_{\delta} & \frac{1}{4}g^2_2 v^2 \cr
\end{matrix}\right),
\label{zpmassmatrix}
\eeqn
where $\alpha=\epsilon c_{\delta}-s_{\delta}$.
The gauge eigenstates of $\mathcal{M'}^2_V$ can be rotated into the corresponding mass eigenstates $(Z',Z,\gamma)$ using the second transformation via the rotation matrix
\beqn
\mathcal{R}=\left(\begin{matrix} c'_W c_{\phi}-s_{\theta}s_{\phi}s'_W & s'_W c_{\phi}+s_{\theta}s_{\phi}c'_W & -c_{\theta}s_{\phi} \cr
c'_W s_{\phi}+s_{\theta}c_{\phi}s'_W & s'_W s_{\phi}-s_{\theta}c_{\phi}c'_W & c_{\theta}c_{\phi} \cr
-c_{\theta} s'_W & c_{\theta} c'_W & s_{\theta} \cr
\end{matrix}\right),
\label{rotmatrix}
\eeqn
with $c'_W(c_{\theta})(c_{\phi})\equiv \cos\theta'_W(\cos\theta)(\cos\phi)$ and $s'_W(s_{\theta})(s_{\phi})\equiv \sin\theta'_W(\sin\theta)(\sin\phi)$, where $\theta'_W$ represents the mixing angle between the new gauge sector and the standard model gauge bosons while the other angles are given by
\begin{equation}
\tan\phi=\alpha, ~~~ \tan\theta=\frac{g_Y}{g_2}c_{\delta}\cos\phi,
\end{equation}
such that $\mathcal{R}^T\mathcal{M'}^2_V\mathcal{R}=\text{diag}(M^2_{Z'},M^2_{Z},0)$.
Defining $M_W=g_2 v/2$, $M_{Z'}\equiv M_+$ and $M_{Z}\equiv M_-$, the angle $\theta'_W$ is given by
\begin{equation}
\tan2\theta'_W\simeq\frac{2\alpha M^2_Z\sin\theta}{M^2_{Z'}-M^2_Z+(M^2_{Z'}+M^2_Z-M^2_W)\alpha^2}.
\end{equation}

%%%%%%%%%%%%%%%%%%%%%%%%%%%%%%%%%%%%%%%%%%%%%%%

\section{Dark matter relic density from freeze-in and freeze-out}\label{sec:DM}

As discussed in Section~\ref{sec:intro}, dark matter in the hidden sector may have couplings with the visible sector which are ultraweak.
Using the analysis of Section~\ref{sec:model}, the lightest particle of the extended model is the hidden sector neutralino $\tilde\xi_1^0$. We assume that the ultraweak particles were not produced in the reheating phase of the early universe.
Further, because of their ultraweak interactions they were never in thermal equilibrium. Thus we assume no relic density for $\tilde\xi_1^0$ at the reheating temperature, i.e., $Y_{\tilde\xi_1^0}=0$ at $T_R$. This is the standard assumption made for
the ultraweak or feeble particles~\cite{Hall:2009bx,Belanger:2018mqt} which we adopt in this study.

For a generic analysis, we denote this particle by $\xi$ and assume it has  a negligible abundance in the early universe. However, since  $\xi$ is the lightest particle in the bath,
 all the heavier $R$-parity odd particles, though ultraweakly coupled to $\xi$, will eventually decay in time to it.
 This implies that the  abundance of $\xi$ will rise as the temperature $T$ drops until the decaying particles run out leading to a saturation in the abundance of $\xi$. For a decaying particle of mass $M$  the dominant production of $\xi$ occurs at $T\sim M$ while the production is
 Boltzmann suppressed for $M>T$.   Below we give an overview of the calculations of the relic density via freeze-in (FI)~\cite{Hall:2009bx,Belanger:2018mqt} then specialize to the specific case where the NLSP is a stop.

For a flat universe, the first Friedman equation reads
\begin{equation}
H^{2}=\left(\frac{\dot{R}}{R}\right)^{2}=\frac{8\pi G}{3}\rho\,,
\label{friedman}
\end{equation}
where $\rho$ is the energy density and $G$ is Newton's gravitational constant. In the radiation dominated universe (for a photon temperature $T\gtrsim 100$ eV), the entropy and energy densities can be written as
\begin{align}
s(T) & =\frac{2\pi^{2}}{45}T^{3}g_{*S}\,, \label{den1}\\
\rho(T) & =\frac{\pi^{2}}{30}T^{4}g_{*}\,,
\label{den2}
\end{align}
where
\begin{align}
g_{*} & =\sum_{i={\rm boson}}g_{i}\left(\frac{T_{i}}{T}\right)^{4}+\frac{7}{8}\sum_{i={\rm fermion}}g_{i}\left(\frac{T_{i}}{T}\right)^{4}\,,\\
g_{*S} & =\sum_{i={\rm boson}}g_{i}\left(\frac{T_{i}}{T}\right)^{3}+\frac{7}{8}\sum_{i={\rm fermion}}g_{i}\left(\frac{T_{i}}{T}\right)^{3}\,,
\end{align}
and $g_{i}$ counts the particle internal degrees of freedom at a temperature $T_i$. Substituting Eq.~(\ref{den2}) into Eq.~(\ref{friedman}) one finds
\begin{equation}
H=\sqrt{\frac{8\pi^{3}}{90}}\frac{\sqrt{g_{*}}}{M_{{\rm pl}}}T^{2}\approx1.66\sqrt{g_{*}}\frac{T^{2}}{M_{{\rm pl}}}\,,
\end{equation}
with $M_{\rm pl}$ being the Planck mass. Using the fact that entropy per comoving volume is conserved, namely $(sR^{3})={\rm const}$, and taking the time derivative one has
\begin{equation}
\frac{{\rm d}s}{{\rm d}t}=-3Hs\,,
\end{equation}
where $H\equiv\dot{R}/R$. Using Eq.~(\ref{den1}), one gets
\begin{equation}
\frac{{\rm d}T}{{\rm d}t}=-\left(\frac{H(T)}{1+\frac{1}{3}\frac{d\ln g_{*S}}{d\ln T}}\right)T.
\label{dt}
\end{equation}
Denoting the quantity in the parentheses of Eq.~(\ref{dt}) $H'(T)$ gives
\begin{equation}
\frac{{\rm d}T}{{\rm {\rm d}}t}=-H'(T)T\,.
\end{equation}
Next  focusing on the reaction $X\leftrightarrows Y+\xi$ where both $X$
and $Y$ are in the thermal bath and $\xi$ is the dark matter particle,  the Boltzmann equation for the number density of $\xi$ reads
\begin{align}
\dot{n}_{\xi}+3Hn_{\xi} & =\int{\rm d}\Pi_{\xi}{\rm d}\Pi_{X}{\rm d}\Pi_{Y}(2\pi)^{4}\delta^{4}(p_{X}-p_{Y}-p_{\xi})\nonumber \\
 & \qquad\times\left[|\overline{\mathcal{M}}|_{X\to Y+\xi}^{2}f_{X}(1\pm f_{Y})(1\pm f_{\xi})-|\overline{\mathcal{M}}|_{Y+\xi\to X}^{2}f_{Y}f_{\xi}(1\pm f_{X})\right]\,,
 \label{bolt}
\end{align}
where ${\rm d}\Pi_{i}=\frac{{\rm d}^{3}p_{i}}{(2\pi)^{3}2E_{i}}$ are phase space elements, $f_{i}$ is the phase space density defined by
\begin{equation}
f_i=\frac{1}{\exp(E_i-\mu_{c_i})/T\pm 1}\,,
\label{density.1}
\end{equation}
where
the plus sign in Eq.~(\ref{bolt}) and in the denominator on the right-hand-side of Eq. (\ref{density.1})
 is for
 bosons and minus for fermions. In Eq.~(\ref{bolt}), $|\overline{\mathcal{M}}|^2$ are summed over initial and final spin and color states. We introduce the fugacity $z$ of the system as $z=z_f e^{\mu_c/T}$ with $\mu_c$ being the chemical potential and $z_f = +1$ for a boson, $-1$ for a fermion and zero for a dark matter particle. The matrix element squared, $|\mathcal{M}|^2$, which enters in the decay width of $X\to Y +\xi$,
   is averaged over initial spin and color states and summed over final spin and color states.
   Thus the decay width of the process $X\to Y+\xi$ is given by
\begin{equation}
\Gamma_X=\frac{1}{2m_{X}}\left(\prod_{i}\frac{{\rm d}^{3}p_{i}}{(2\pi)^{3}}\frac{1}{2E_{i}}\right)|\mathcal{M}|_{X\to Y+\xi}^{2}(2\pi)^{4}\delta^{4}(p_{X}-p_{Y}-p_{\xi})\,.
\label{width}
\end{equation}
 Assuming the initial $\xi$ abundance is zero, i.e., $f_{\xi}=0$, the term corresponding to $Y+\xi\to X$ in Eq.~(\ref{bolt}) vanishes. Further, we set $1+f_Y\sim 1$ which reduces Eq.~(\ref{bolt}) to the following
 \begin{align}
\dot{n}_{\xi}+3Hn_{\xi} & =\int{\rm d}\Pi_{\xi}{\rm d}\Pi_{X}{\rm d}\Pi_{Y}(2\pi)^{4}\delta^{4}(p_{X}-p_{Y}-p_{\xi})
 |\overline{\mathcal{M}}|_{X\to Y+\xi}^{2}f_{X}.
  \label{bolt-1}
\end{align}
Noting that $|\overline{\mathcal{M}}|^2= g_{X} |\mathcal{M}|^2$ and using Eq.~(\ref{width}), we can write Eq.~(\ref{bolt-1}) so that
 \begin{equation}
\dot{n}_{\xi}+3Hn_{\xi} = \frac{m_X^2 g_{X} \Gamma_X }{ 2\pi^2} T K_1(x_X),
   \label{bolt-3}
\end{equation}
where
\begin{align}
K_1(x_X)=  \int {\rm d}u~  x_X (u^2-1)^{1/2} e^{-u x_X},
\label{bessel}
\end{align}
is the Bessel function of the second kind and degree one. Now note that defining $Y_{\xi}= n_{\xi}/s$, one gets
\begin{align}
Y_{\xi}\simeq \int  \frac{m_X^2 g_{X} \Gamma_X }{ 2\pi^2s} T K_1(x_X)~{\rm d}t.
\label{Yx1}
\end{align}
Next we use the relation between time and temperature which is ${\rm d}t= -\dfrac{{\rm d}T} {H(T) T}$. Using this in Eq.~(\ref{Yx1}) we get,
 \begin{align}
Y_{\xi}\simeq \frac{g_{X}}{2\pi^2}\Gamma_X m^2_X \int_{T_{\rm min}}^{T_{\rm max}}   \frac{{\rm d}T}{s(T)H(T)} K_1(x_X).
\label{Yx2}
\end{align}

In the numerical analysis we use the more exact form of $Y_{\xi}$ given  by
\begin{equation}
Y_{\xi}=\frac{g_{X}|z_X|}{2\pi^2}\Gamma_X m^2_X \int_{T_0}^{T_R}\frac{{\rm d}T}{H'(T)s(T)}K'_1(x_X,x_{\xi},x_Y,z_X,z_{\xi},z_Y),
\label{yield}
\end{equation}
where $T_0$ is the current temperature and $T_R$ is the reheating temperature and we have defined $K'_1$ as the generalized Bessel function of the second kind of degree one given by
\begin{equation}
K'_1(x_X,x_{\xi},x_Y,z_X,z_{\xi},z_Y)=x_X\int_1^{\infty}\frac{du\sqrt{u^2-1}e^{-x_X u}}{1-z_X e^{-x_X u}}S(x_X\sqrt{u^2-1},x_X,x_{\xi},x_Y,z_{\xi},z_Y),
\label{genBessel}
\end{equation}
with the function $S$ defined in~\cite{Belanger:2018mqt} as
\begin{equation}
S(p_X/T,x_X,x_{\xi},x_Y,z_{\xi},z_Y)=\frac{1+\frac{m_X T}{2p_X p_{\xi,Y}}\log\left[\frac{(1-z_Y e^{-E_Y(1)/T})(1-z_{\xi} e^{-E_{\xi}(-1)/T})}{(1-z_{\xi} e^{-E_{\xi}(1)/T})(1-z_Y e^{-E_Y(-1)/T})}\right]}{1-z_Yz_{\xi}e^{-E_X/T}},
\end{equation}
where neglecting the effect of the chemical potential, i.e. setting $z_Y$ and $z_{\xi}$ to zero, $S\rightarrow 1$ and so Eq.~(\ref{genBessel}) reduces to Eq.~(\ref{bessel}). The function $K_1'$ which takes six arguments corresponding to values of
$x_{X,\xi,Y}$ where $x=m/T$ and by the  fugacity parameters $z_{X,\xi,Y}$ is evaluated using \code{micrOMEGAs5.0} routines.

For our benchmarks, the NLSP is the stop and so one of the reactions contributing to dark matter production via FI is $\tilde t\to \tilde\xi^0_1 t$. Taking $z_{\tilde t}=+1$, $z_{t}=-1$ and $z_{\tilde\xi^0_1}=0$, Eq.~(\ref{yield}) takes the form
 \begin{equation}
Y_{\tilde\xi^0_1}=\frac{g_{\tilde t}}{2\pi^2}\Gamma_{\tilde t}~ m^2_{\tilde t} \int_{T_0}^{T_R}\frac{{\rm d}T}{H'(T)s(T)}K'_1(x_{\tilde t},x_{\tilde\xi^0_1},x_t,1,0,-1),
\label{yieldf}
\end{equation}
where $g_{\tilde t}=6$. The integral of Eq.~(\ref{yieldf}) is evaluated numerically and using
\begin{equation}
\Omega h^2=\frac{m Y s_0 h^2}{\rho_c},
\label{reliceq}
\end{equation}
we calculate the FI contribution to the relic density, i.e., $(\Omega h^2)_{\rm FI}$. In Eq.~(\ref{reliceq}), $s_0$ is today's entropy density, $\rho_c$ is the critical density and $h=0.678$.

Next we use the benchmarks of Table~\ref{tab1} of Section~\ref{sec:implement} to exhibit in the left panel of
Fig.~\ref{figa} the comoving number density of the hidden sector neutralino and the stop as a function of $x=m_{\tilde\xi^0_1}/T$ for the freeze-in case. Here one finds that at small $x$, i.e. at high temperatures the abundance of $\tilde\xi^0_1$ is negligible as expected
and starts to grow as the temperature drops until reaching its saturation value at $x\sim 3-5$ while the abundance of the stop decreases with $x$ due to the slow decay of the stop into the hidden sector neutralino. The four curves correspond to four of our ten benchmarks of
Table \ref{tab1} and the plot is only drawn for the abundance obtained by the decay of a stop. To understand the order of those curves, we note  that the comoving number density at saturation is $Y^{\rm max}_{\tilde\xi^0_1}\propto \Gamma_{\tilde t}/m^2_{\tilde t}$ and benchmarks (a), (c), (e) and (g) have an increasing stop mass  which explains the order of the curves in the left panel of Fig.~\ref{figa}.

\begin{figure}[H]
 \centering
   \includegraphics[width=0.48\textwidth]{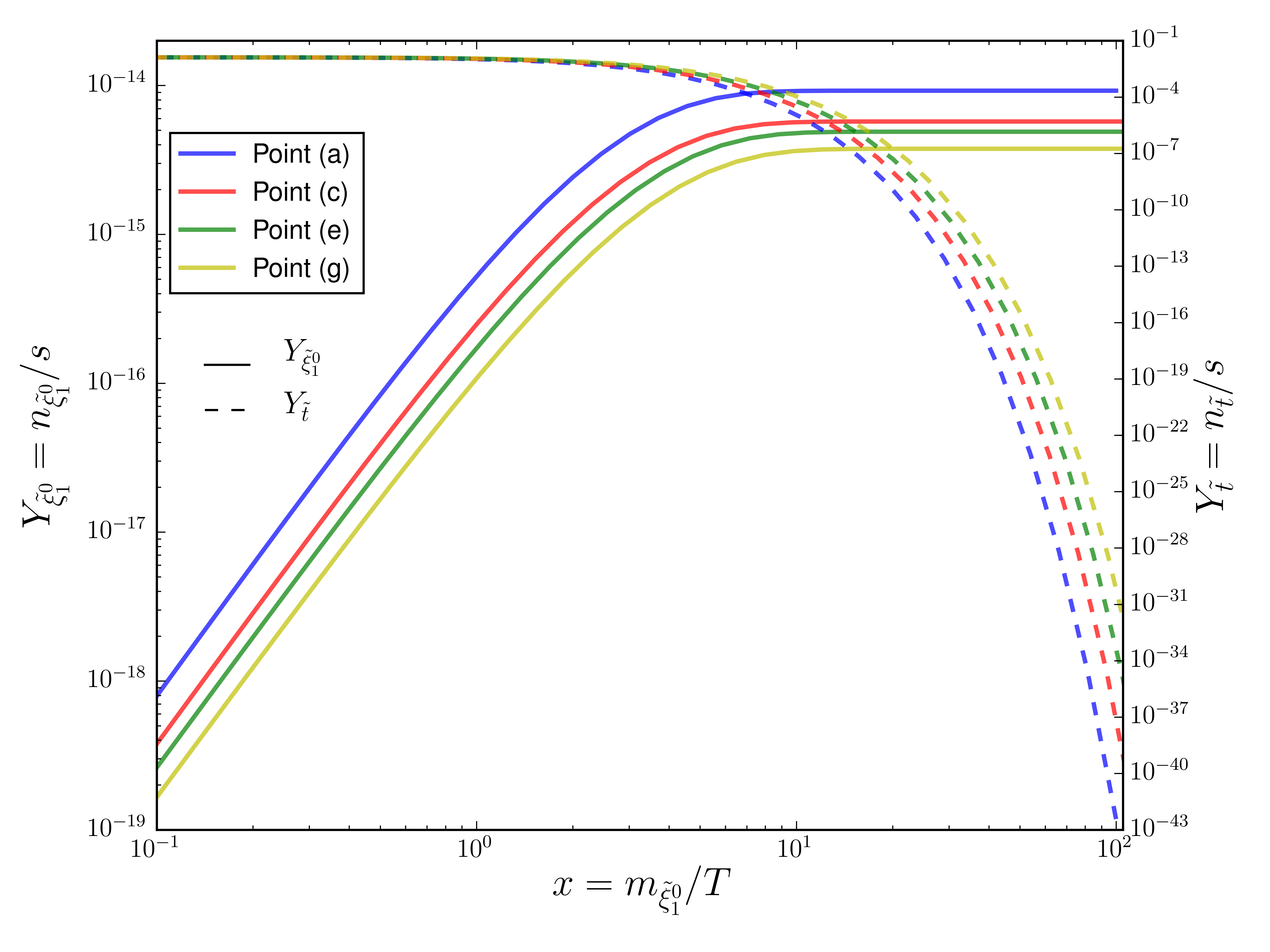}
   \includegraphics[width=0.48\textwidth]{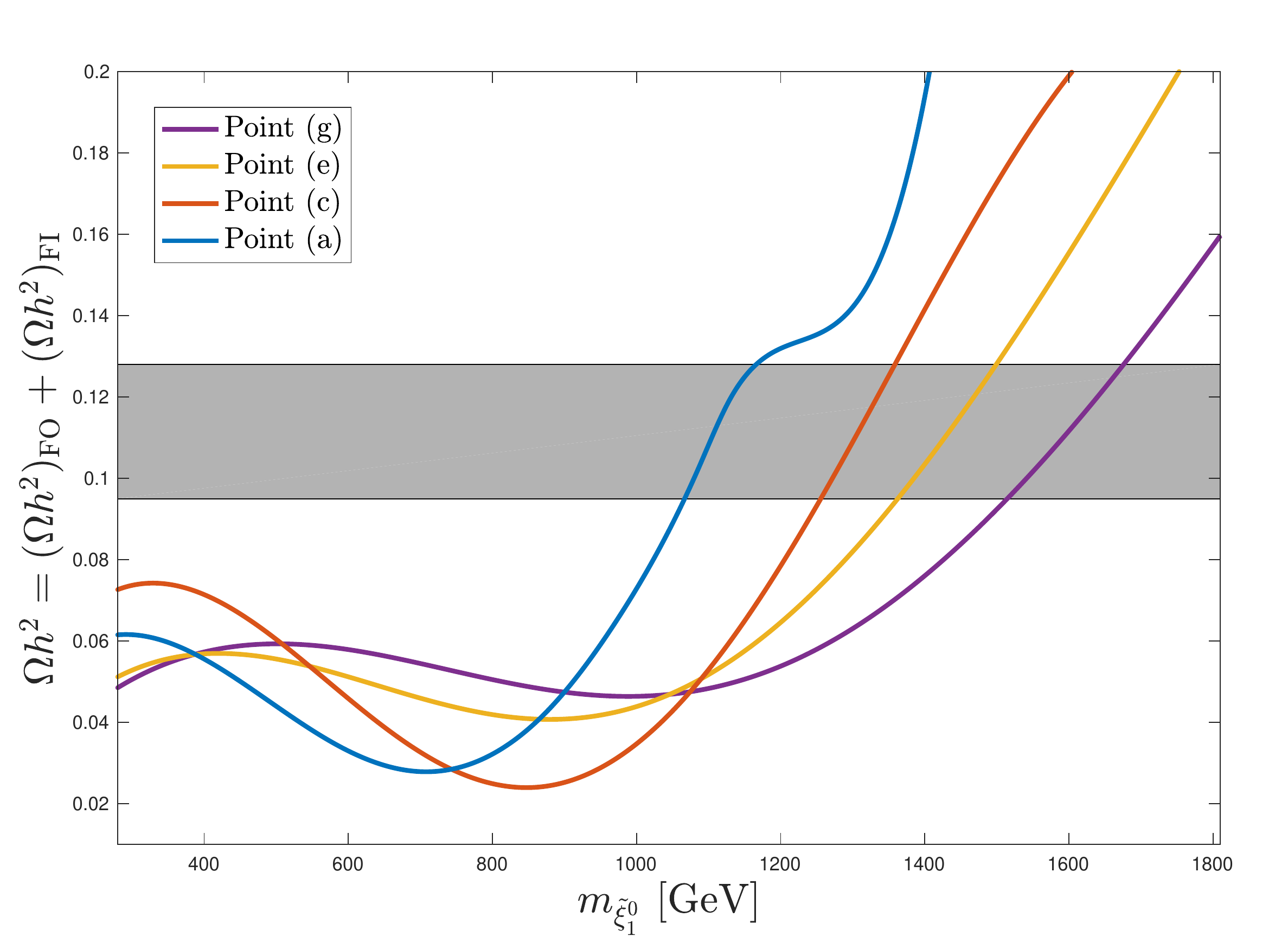}
   \caption{Left panel: a plot of the comoving number density $Y_{\tilde\xi^0_1}$ and $Y_{\tilde t}$ versus $x$ for four illustrative benchmarks (a), (c), (e) and (g) of Table~\ref{tab1} for the freeze-in situation. Right panel: a plot of the total relic density from FI and FO versus the dark matter mass for the same benchmarks. The mass range is obtained by varying $M_X$ and keeping the rest of the input parameters the same. The grey patch shows the allowed region of the relic density taking theoretical uncertainties into account.}
	\label{figa}
\end{figure}

The second contribution to the relic density is due to the freeze-out processes. However, the freeze-out contribution is not from the $\tilde\xi^0_1\tilde\xi^0_1$  annihilation  which, as discussed earlier, is assumed negligible. Rather, it arises from the freeze-out of the stops
 which are in thermal equilibrium with the bath in the early universe. Once out of equilibrium, the stops then decay to $\tilde\xi^0_1$ to make up the freeze-out (FO) portion of the relic density. Using the standard FO considerations, one can determine the relic density of the stops, $(\Omega h^2)^{\tilde t}_{\rm FO}$,  using \code{micrOMEGAs}  and the relic density of $\tilde\xi^0_1$ is given by
\begin{equation}
(\Omega h^2)_{\rm FO}=\frac{m_{\tilde\xi^0_1}}{m_{\tilde t}}(\Omega h^2)^{\tilde t}_{\rm FO}\,.
\end{equation}
The total relic density as given in Table~\ref{tab2} is then
\begin{equation}
\Omega h^2=(\Omega h^2)_{\rm FO}+(\Omega h^2)_{\rm FI}\,.
\label{totrelic}
\end{equation}
Thus the total relic density receives  contributions from both freeze-out and freeze-in mechanisms and is consistent with the current value of the dark matter relic density as measured by the Planck experiment~\cite{Aghanim:2018eyx}
\begin{equation}
\Omega h^2=0.1198\pm 0.0012,
\label{relic}
\end{equation}
for all the benchmarks of Table~\ref{tab1}.

The total relic density of Eq.~(\ref{totrelic}) is plotted against the dark matter mass in the right panel of Fig.~\ref{figa} for four benchmarks (a), (c), (e) and (g) of Table~\ref{tab1}. The grey patch shows the acceptable region of dark matter relic density taking into account the theoretical uncertainties. The FO contribution to the relic density has a linear dependence on $m_{\tilde\xi^0_1}$ and so this non-linear variation in $\Omega h^2$ is driven by the FI contribution which is proportional to $m_{\tilde\xi^0_1}\Gamma_{\tilde t}/m^2_{\tilde t}$. For a fixed stop mass, as the dark matter mass increases (approaching the stop mass), the ratio $m_{\tilde\xi^0_1}/m^2_{\tilde t}$ becomes larger and competes with the falling decay width causing a steady rise in the FI relic density. However, for $m_{\tilde\xi^0_1}$ smaller than a certain threshold, the decay width begins to compete with the decreasing $m_{\tilde\xi^0_1}/m^2_{\tilde t}$ eventually leading to an increase in the FI relic density even for small dark matter masses. This trend can be clearly seen in the right panel of Fig.~\ref{figa}.

Before concluding this section, it is worth noting that the freeze-in contribution is most relevant when the stop annihilation cross-section is the largest. This general trend can be seen from Table~\ref{tab2} where the FO relic increases with increasing stop mass while the opposite happens for the FI relic density. For heavier stops, the annihilation cross-section drops and with this the FO relic density increases. As a result, the FI contribution decreases which can also be seen from its inverse dependence on $m^2_{\tilde t}$.
We note that in the above we have not taken into account the effect of CP phases on the soft parameters in the MSSM analysis. Such phases, however, are likely to
affect the analysis to order a few percent (see, e.g.,~\cite{Gomez:2006uv}) and not drastically change the conclusions of the analysis given here.

\section{Model implementation and long-lived stop}\label{sec:implement}

For a phenomenological study of the model described in Section~\ref{sec:model}, we use the mathematica package \code{SARAH-4.14}~\cite{Staub:2013tta,Staub:2015kfa} to generate model files for the spectrum generator \code{SPheno-4.0.3}~\cite{Porod:2003um,Porod:2011nf} which runs the renormalization group equations (RGE) starting from a high scale input to produce the sparticle masses and calculate their decay widths.  \code{SARAH} also generates \code{CalcHep/CompHep}~\cite{Pukhov:2004ca,Boos:1994xb} files used by \code{micrOMEGAs-5.0.4}~\cite{Belanger:2014vza} to determine the dark matter (DM) relic density via the freeze-out and freeze-in routines and \code{UFO} files~\cite{Degrande:2011ua} which are input to \code{MadGraph5}~\cite{Alwall:2014hca}.

The input parameters of the $U(1)_X$-extended MSSM/SUGRA~\cite{msugra,Nath:2016qzm} are of the usual non-universal SUGRA model with additional parameters as below (all at the GUT scale)
\begin{equation}
m_0, ~~A_0, ~~ m_1, ~~ m_2, ~~ m_3, ~~M_1, ~~m_X, ~~\delta, ~~\tan\beta, ~~\text{sgn}(\mu).
\label{sugra}
\end{equation}
where $m_0, ~A_0, ~m_1, ~m_2, ~m_3, ~\tan\beta$ and $\text{sgn}(\mu)$ are the soft parameters in the MSSM sector as defined earlier.
The parameters $M_2$ and $M_{XY}$ are set to zero at the GUT scale. However, those parameters acquire a tiny value at the electroweak scale due to RGE running. In scanning the parameter space of the model we accept points satisfying the Higgs boson mass and DM relic density constraints. Taking theoretical uncertainties into consideration, the constraint of the Higgs mass is at $125\pm 2$ GeV while the relic density is in the range 0.110$-$0.128 and both constitute the first level of constraints. More requirements coming from LHC data and cosmology are imposed thereafter (discussed later). We select ten benchmarks satisfying all the previous constraints and are displayed in Table~\ref{tab1}.

\begin{table}[H]
\begin{center}
\begin{tabulary}{1.00\textwidth}{l|CCCCCCCCC}
\hline\hline\rule{0pt}{3ex}
Model & $m_0$ & $A_0$ & $m_1$ & $m_2$ & $m_3$ & $M_1$ & $m_X$ & $\tan\beta$ & $\delta$ \\
\hline\rule{0pt}{3ex}
\!\!(a) & 2632 & -6455 & 3150 & 2100 & 1450 & 1305 & 380 & 20 & $1.02\times10^{-11}$   \\
(b) & 4122 & -7760 & 3363 & 2622 & 1165 & 1400 & 380 & 15 & $1.00\times10^{-11}$ \\
(c) & 2106 & -4366 & 3756 & 2080 & 1263 & 1533 & 380 & 18 & $1.03\times10^{-11}$ \\
(d) & 5042 & -9280 & 4163 & 3044 & 1206 & 1522 & 450 & 10 & $1.10\times10^{-11}$ \\
(e) & 3382 & -7593 & 4046 & 2746 & 1695 & 1720 & 510 & 23 & $8.80\times10^{-12}$ \\
(f) & 4825 & -7565 & 4551 & 3862 & 1097 & 1885 & 805 & 13 & $9.50\times10^{-12}$ \\
(g) & 3851 & -6784 & 4950 & 3277 & 1426 & 1973 & 712 & 25 & $9.00\times10^{-12}$ \\
(h) & 5624 & -9330 & 7532 & 5250 & 1434 & 2105 & 850 & 8 & $1.15\times10^{-11}$ \\
(i) & 6158 & -10265 & 5000 & 4895 & 1303 & 1944 & 586 & 28 & $7.00\times10^{-12}$ \\
(j) & 6638 & -11055 & 6532 & 5200 & 1507 & 2036 & 638 & 5 & $8.50\times10^{-12}$ \\
\hline
\end{tabulary}\end{center}
\caption{Input parameters for the benchmarks used in this analysis. Here $M_2=M_{XY}=0$ at the GUT scale. All masses are in GeV.}
\label{tab1}
\end{table}

The search for promptly decaying stops at the LHC targets non-leptonic (high $p_T$ jets along with large missing transverse energy) and leptonic final states. For non-compressed spectra, the latest searches with the most stringent constraints on the stop mass are from ATLAS~\cite{Aaboud:2017ayj} where a stop mass up to 1 TeV is excluded for an LSP mass less than 160 GeV and from CMS~\cite{Sirunyan:2019xwh} with an exclusion limit reaching 1.2 TeV for an LSP mass less than $\sim 400$ GeV using 137$\ifb$. For compressed spectra, the latest search from ATLAS uses 139$\ifb$ of data and excludes stops up to 720 GeV with an LSP up to 580 GeV~\cite{ATLAS:2019oho}. Remarkably, stronger constraints on stop masses come from searches of long-lived stops at the LHC where a stop is considered stable over detector length. Thus ATLAS excludes stops up to $\sim 1.3$ TeV~\cite{Aaboud:2019trc} while CMS has a weaker exclusion limit at $\sim 1$ TeV~\cite{Khachatryan:2016sfv}. Experimental collaborations search for long-lived stops as part of composite objects called $R$-hadrons which form after a stop hadronizes. $R$-hadrons, which is a generic name for stop or gluino $R$-hadrons, have been studied a lot in the experimental community~\cite{ATLAS:2019duq,Aad:2013gva,Mehlhase:2013wva,Aad:2012pra} and less from the theory/phenomenology standpoint. In the latter, long-lived stops which are degenerate with the neutralino LSP~\cite{Johansen:2010ac,Kim:2011sv} or with the gravitino LSP~\cite{Diaz-Cruz:2019xjb,Covi:2014fba} have been studied in the MSSM while considering visible sector dark matter candidates. In this work we do not require a small mass gap between the stop and the DM candidate as the tiny stop decay width arises only due to the very weak couplings between the visible and hidden sectors.

In Table~\ref{tab2} below we present the stop, gluino and electroweakino masses for our ten benchmarks of Table~\ref{tab1}. The stop mass ranges from 1.4 TeV to 2.3 TeV which satisfy the exclusion limits from ATLAS and CMS as described above. Further, all gluinos have masses greater than 2.5 TeV and electroweakinos are in the TeV range.

\begin{table}[H]
\begin{center}
\begin{tabulary}{1.3\textwidth}{l|CCCCCCCCCCC}
\hline\hline\rule{0pt}{3ex}
Model  & $h^0$ & $\mu$ & $\tilde\chi_1^0$ & $\tilde\chi_1^\pm$ &  $\tilde{\xi}^0_1$ & $\tilde t$ & $\tilde g$ & $(\Omega h^2)_{\rm FO}$ & $ (\Omega h^2)_{\rm FI}$ & $\Omega h^2$ & $\tau_0$ \\
\hline\rule{0pt}{3ex}
\!\!(a) & 124.2 & 3122 & 1416 & 1759 & 1129 & 1409 & 3218 & 0.044 & 0.076 & 0.119 & 0.79 \\
(b) & 125.5 & 3168 & 1529 & 2218 & 1223 & 1502 & 2709 & 0.046 & 0.070 & 0.116 & 0.81 \\
(c) & 124.4 & 2324 & 1678 & 1727 & 1355 & 1618 & 2821 & 0.038 & 0.089 & 0.127 & 0.97 \\
(d) & 125.6 & 3665 & 1907 & 2587 & 1314 & 1702 & 2817 & 0.047 & 0.065 & 0.112 & 0.43 \\
(e) & 125.5 & 3556 & 1836 & 2310 & 1484 & 1804 & 3737 & 0.065 & 0.059 & 0.124 & 0.91 \\
(f) & 125.4 & 2763 & 2085 & 2773 & 1525 & 1903 & 2575 & 0.065 & 0.044 & 0.110 & 0.84 \\
(g) & 125.8 & 2900 & 2254 & 2737 & 1649 & 2005 & 3224 & 0.073 & 0.050 & 0.122 & 0.96 \\
(h) & 125.6 & 3513 & 3461 & 3519 & 1722 & 2102 & 3284 & 0.081 & 0.040 & 0.121 & 0.92 \\
(i) & 126.8 & 3444 & 2316 & 3465 & 1673 & 2201 & 3033 & 0.085 & 0.030 & 0.115 & 0.66 \\
(j) & 123.7 & 4454 & 3034 & 4360 & 1742 & 2304 & 3460 & 0.088 & 0.031 & 0.119 & 0.55 \\
\hline
\end{tabulary}\end{center}
\caption{Display of the Higgs boson ($h^0$) mass, the $\mu$ parameter, the stop mass,  the relevant electroweak gaugino masses, and the relic density for the benchmarks  of Table~\ref{tab1} computed at the electroweak scale. The lifetime, $\tau_0$ (in s) of the long-lived stop is also shown. All masses are in GeV. }
\label{tab2}
\end{table}

The last column in Table~\ref{tab2} shows the proper lifetime of a long-lived stop and all of which are less than one second. This is in agreement with the cosmological constraint from Big Bang Nucleosynthesis (BBN) which requires the lifetime of long-lived particles to be $\mathcal{O}$(1$-$10) seconds so that the BBN's prediction of light nuclei abundance in the early universe is not disrupted~\cite{Kawasaki:2004qu,Kawasaki:2017bqm}.
In Table~\ref{tab2}  we also display the relative contributions from freeze-in given by $(\Omega h^2)_{\rm FI}$ and freeze-out given by  $(\Omega h^2)_{\rm FO}$.
For model point (c), the freeze-in contribution to the total relic density is about 70\% but is only about 26\% for model point (j). Typically for relatively small stop masses with
relatively large stop annihilation cross-sections, the freeze-in relic density tends to dominate the freeze-out part. However, for relatively larger stop masses with relatively
small annihilation cross-sections the freeze-out part tends to dominate the freeze-in part. Importantly, the freeze-out contribution is found never to be negligible  relative to the
freeze-in part and thus the freeze-in alone is not sufficient for the relic density analysis. This is the case for the entire set of model points considered in Table~\ref{tab1} and the pattern described in shown more explicitly in Fig.~\ref{figb}.  The inversion in the FI and FO contributions to the relic density as a function of the stop mass can be clearly seen as described before while the total relic density lies entirely in the acceptable region (grey patch).
In the analysis of Table~\ref{tab2}  the model points satisfy the relic density constraint consistent with  Planck~\cite{Aghanim:2018eyx} only for the sum of freeze-in and freeze-out.

\begin{figure}[H]
 \centering
   \includegraphics[width=0.6\textwidth]{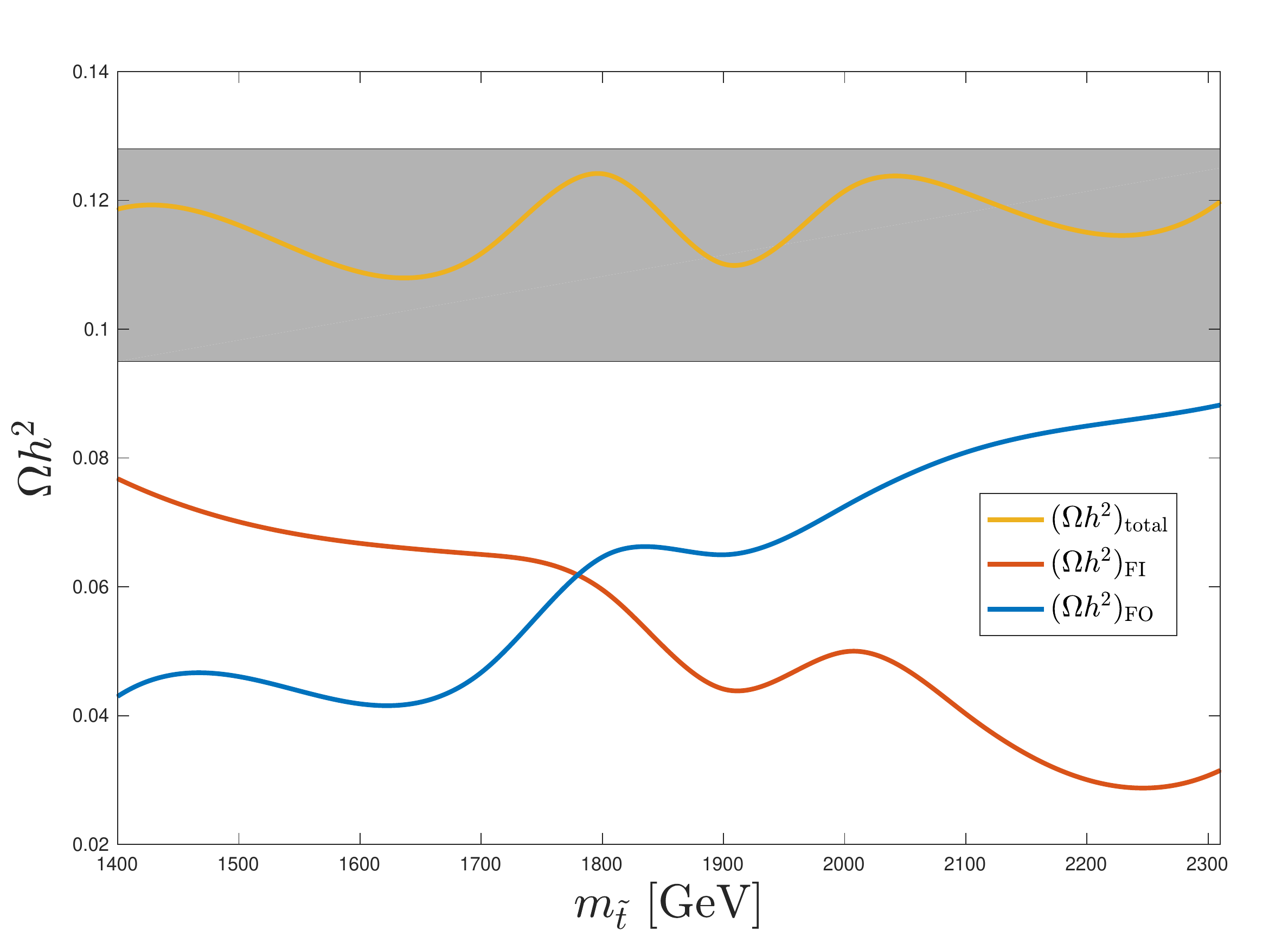}
   \caption{A plot of the relic density versus the stop mass for all the benchmarks of Table~\ref{tab1}. The FI and FO contributions are shown along with their sum which lies inside the grey patch defined in Fig.~\ref{figa}.}
	\label{figb}
\end{figure}

%%%%
 Before we conclude this section we give a brief account of stop $R$-hadrons and their properties.
Long-lived stops (with a decay width $\lesssim 0.2$ GeV) immediately hadronize forming color-neutral $R$-hadrons, $R_{\tilde t}$, which can be thought of as a stop surrounded by a ``cloud" of light quarks. Around 93\% of $R_{\tilde t}$ formed are $R$-mesons $\tilde t\bar{q}$ and the rest are $R$-baryons $\tilde t qq$. Interactions of $R$-hadrons with detector material are largely understood as they mainly arise due to light quarks since stops have a small interaction cross-section. For this reason, energy deposited in the calorimeters is small (typically less than 10 GeV).
As a result of  interactions between the R-hadrons and detector material,
%this interaction
most of the $R_{\tilde t}$ transform from mesons to baryons. This transition leads to charge flipping where an $R$-hadron can go from being electrically charged to neutral and vice-versa. On the average, almost half of the $R$-hadrons end up flipping sign~\cite{Hohansen:2007kb} as they travel the detector length. Since the stop parton of $R_{\tilde t}$ is electrically charged, more than half ($\sim 57$\%) of $R$-hadrons are formed with an electric charge~\cite{ATLAS:2019duq} and will, therefore, leave a track in the inner detector tracker (ID) and in the muon spectrometer (MS). Due to the charge flipping property, tracks may suddenly disappear or appear which is a feature used by experimental collaborations to look for $R$-hadrons. A track in the ID may have no corresponding track in the MS and vice-versa. An $R$-hadron composed of an anti-stop is unlikely to transition from an anti-meson state to an anti-baryon. However, if it happens, the anti-baryonic state will annihilate back to a anti-mesonic state as it interacts with the detector material.

\section{Stop pair production at the LHC}\label{sec:production}

In the MSSM, the stop mass receives contributions from terms in the superpotential and from soft SUSY breaking terms. The mass-squared matrix for stop quarks defined in the gauge eigenstate basis $(\tilde t_L,\tilde t_R)$ is given by
\begin{equation}
\mathcal{L}_{\tilde t}=-\left(\begin{matrix}\tilde t^*_L & \tilde t^*_R \end{matrix}\right)M^2_{\tilde t}\left(\begin{matrix}\tilde t_L \\ \tilde t_R \end{matrix}\right),
\end{equation}
where
\begin{equation}
M^2_{\tilde t}=\left(\begin{matrix} m^2_{\tilde t_R} & m_t(A_t-\mu\cot\beta) \\
m_t(A_t-\mu\cot\beta) & m^2_{\tilde t_L}\end{matrix}\right).
\label{stopmat}
\end{equation}
Each of the diagonal entries of this hermitian matrix is a sum of the relevant soft SUSY breaking term, a $D$ term and the top mass-squared. The off-diagonal entries are given in terms of the top trilinear coupling $A_t$, the top mass $m_t$ and $\mu$ and $\tan\beta$ as defined in Section~\ref{sec:implement}. For a particular choice of $A_0$ at the GUT scale, the obtained value of $A_t$ at the electroweak scale can be large enough to generate a considerable mass splitting between the two top mass eigenstates, $\tilde t_1$ and $\tilde t_2$ obtained by rotating the gauge eigenstates. The lightest of those states is $\tilde t_1$ which we have been simply denoting as $\tilde t$ throughout.

The production of stops at the LHC may proceed directly or indirectly following the decay of heavier strongly interacting particles. For instance, the production of gluinos $\tilde{g}$ may be followed by the decay $\tilde g\rightarrow t^{(*)}\tilde t$ which will be the source of stops. From Table~\ref{tab2}, gluinos are more than a TeV heavier than stops, so the production cross-section of a gluino pair is suppressed in comparison to a stop pair production. Hence it suffices to consider direct stop pair production for our study. The production of a stop-antistop pair proceeds via the leading partonic processes
\begin{align}
gg&\rightarrow \tilde t\tilde t^*, \nonumber \\
q\bar{q}&\rightarrow \tilde t\tilde t^*,
\end{align}
with respective cross-sections at leading order (LO) given by~\cite{Beenakker:1997ut}
\begin{align}
\hat{\sigma}_{\rm LO}(gg\to \tilde t\tilde t^*)&=\frac{\alpha_s\pi}{s}\left[\beta_0\left(\frac{5}{48}+\frac{31m^2_{\tilde t}}{24s}\right)+\left(\frac{2m^2_{\tilde t}}{3s}+\frac{m^4_{\tilde t}}{6s^2}\right)\log\left(\frac{1-\beta_0}{1+\beta_0}\right)\right], \label{gg} \\
\hat{\sigma}_{\rm LO}(q\bar{q}\to \tilde t\tilde t^*)&=\frac{2\alpha_s\pi}{27s}\beta^3_0,
\label{qq}
\end{align}
where $\alpha_s$ is the strong coupling constant, $\sqrt{s}$ is the invariant center of mass energy and $\beta_0=\sqrt{1-4m^2_{\tilde t}/s}$. From Eqs.~(\ref{gg}) and~(\ref{qq}) one finds that the gluon fusion process is the dominant one. Stop-antistop cross-section is known at next-to-leading order (NLO)~\cite{Beenakker:1997ut}, at NLO with threshold resummation of next-to-leading logarithm (NLO+NLL)~\cite{Beenakker:2010nq,Borschensky:2014cia} and at NNLO+NNLL~\cite{Beenakker:2016gmf}. We calculate the stop-antistop pair production cross-section using \code{Prospino2}~\cite{Beenakker:1996ed,Beenakker:1999xh} at NLO in QCD and at NLO+NLL with the help of \code{NLL-fast}~\cite{Beenakker:2015rna} at 14 TeV and at 27 TeV using the CTEQ5 PDF set~\cite{Lai:1999wy}. The NLO+NLL cross-sections are $\sim 5\%-8\%$ more than the NLO ones at 14 TeV while the change is less significant at 27 TeV with only a $\sim 2\%-4\%$ increase. Note that stop-antistop cross-sections at NNLO+NNLL are only available in \code{NLL-fast} at 13 TeV.
Same sign stop pair production cross-section is calculated at LO using \code{MadGraph5}. The results are presented in Table~\ref{tab3}.
Due to the smallness of the gauge kinetic and mass mixing coefficients, the contributions from the hidden sector to the production cross-section is negligible and so one can use the MSSM to calculate production cross-sections.

\begin{table}[H]
\begin{center}
\begin{tabulary}{1.2\textwidth}{l|cc|cc}
\hline\hline\rule{0pt}{3ex}
Model  & \multicolumn{2}{c}{$\sigma_{\rm NLO+NLL}(pp\rightarrow \tilde t\,\tilde t^*)$} & \multicolumn{2}{c}{$\sigma_{\rm LO}(pp\rightarrow \tilde t\,\tilde t)$} \\
&  &  &  &  \\
  & 14 TeV & 27 TeV & 14 TeV & 27 TeV  \\
\hline\rule{0pt}{3ex}
\!\!(a) & 0.654 & 13.5 & 0.092 & 1.190 \\
(b) & 0.387 & 9.03  & 0.060 &  0.840 \\
(c) & 0.197 & 5.56 & 0.033 & 0.550  \\
(d) & 0.129 & 4.00 &  0.021&  0.412 \\
(e) & 0.075 & 2.69 & 0.013 & 0.290  \\
(f) & 0.046 & 1.89 & 0.008 &  0.214 \\
(g) & 0.029 & 1.29 & 0.005 & 0.155  \\
(h) & 0.018 & 0.92 & 0.003 & 0.115  \\
(i) & 0.011 & 0.66 & 0.002 & 0.085  \\
(j) & 0.006 & 0.47 & 0.001 &  0.063 \\
\hline
\end{tabulary}
\end{center}
\caption{The NLO+NLL production cross-sections, in fb, of a stop-antistop pair, $\tilde t\,\tilde t^*$ (second and third columns), and the LO cross-sections, in fb, of a stop pair (fourth and fifth columns) at $\sqrt{s}=14$ TeV and at $\sqrt{s}=27$ TeV for benchmarks of Table~\ref{tab1}.}
\label{tab3}
\end{table}

\section{Signal and background simulation and event preselection}\label{sec:simulation}

Our signal consists of long-lived heavy stops traversing the detector at a low speed. In the muon spectrometer (MS) this particle will look like a heavy muon with a large transverse momentum $p_T$. Therefore the main SM backgrounds are processes resulting in muons along with non-physical backgrounds consisting of mismeasurements (of the muon velocity, as an example) and other detector effects. Hence the largest contributors to the physical SM backgrounds are $W/Z/\gamma^*+$ jets, diboson production, single top, $t\bar{t}$ and $t+W/Z$. The signal and background events are simulated at LO with \code{MadGraph5\_aMC@NLO-2.6.3} interfaced to \code{LHAPDF}~\cite{Buckley:2014ana} using the NNPDF30LO PDF set. The cross-sections are then scaled to their NLO values at 14 TeV and at 27 TeV. The resulting files are passed to \code{PYTHIA8}~\cite{Sjostrand:2014zea} for showering and hadronization. For the SM backgrounds, a five-flavour MLM matching~\cite{Mangano:2006rw} is performed on the samples in order to avoid double counting of jets. Jets are clustered with \code{FastJet}~\cite{Cacciari:2011ma} using the anti-$k_t$ algorithm~\cite{Cacciari:2008gp} with jet radius $R=0.4$. For the signal, \code{PYTHIA8} simulates the hadronization of the long-lived stops into $R$-hadrons. Detector simulation and event reconstruction is handled by \code{DELPHES-3.4.2}~\cite{deFavereau:2013fsa} using the beta card for HL-LHC and HE-LHC studies. The analysis of the resulting event files and cut implementation is carried out with \code{ROOT 6}~\cite{Antcheva:2011zz}.

As explained earlier, $R$-hadrons undergo charge-flipping as they traverse the detector length while interacting with the detector material. Thus it is very likely that a visible $R$-hadron track can be detected in the inner tracker with no corresponding track reconstructed in the MS and vice-versa. Unlike \code{GEANT4}~\cite{Agostinelli:2002hh,Mackeprang:2009ad}, the fast detector simulator \code{DELPHES} does not handle such a scenario so we opt to carry out the analysis at the ID level whereby we focus on identifying muons and $R$-hadrons solely using information from the inner tracker of our generic detector.  The signal region (SR) will be called ``ID-only". For this SR, some preselection criteria are in order. In the detector, $R$-hadrons will look like slow moving muons with large transverse momentum $p_T$. Events are selected by identifying muons/$R$-hadrons tracks which are central and have large $p_T$, i.e. $|\eta|<2.4$ rad and $p_T>150$ GeV. An electron veto is applied along with a $Z$ veto which means that events whose reconstructed dimuon mass is within 10 GeV of the $Z$ pole mass are rejected.

\section{Selection criteria and results}\label{sec:results}

Following the preselection criteria mentioned in the previous section, additional cuts are applied to enhance the signal over the SM background. The main jet activity in the signal comes from initial and final state radiation (ISR and FSR) while the SM backgrounds include, along with ISR and FSR, hard jets at generator level. Large missing transverse energy $E^{\rm miss}_T$ arises due to ISR boosting the $R$-hadron system thus creating a momentum imbalance which adds to the $E^{\rm miss}_T$ of the event. The minimum missing transverse energy in each event must meet the trigger requirement of 90$-$120 GeV. To distinguish a candidate track from a possible high $p_T$ jet faking it, we impose a minimum cut on the spatial separation between a track and the leading jet in an event, $\Delta R(\text{track},\text{jet}_1)$. Another important kinematic variable is the speed $\beta_s=p/E$ of a muon/$R$-hadron which must be greater than 0.6 so that an $R$-hadron can be associated with the same bunch crossing and pass the trigger requirement. Much slower $R$-hadrons do not make it in time to be recorded as an interesting physics event. Muons and SM hadrons mostly have $\beta_s\sim 1$ and all events exhibiting $\beta_s<1$ are due to mismeasurements and must be accounted for. We list the kinematic variables and their cut values in Table~\ref{tab4} below.

\begin{table}[H]
	\begin{center}
	\begin{tabulary}{1.0\linewidth}{l|cccc}
    \hline\hline
	& \multicolumn{4}{c}{``ID-only'' SR} \\
    \cline{2-5}
   Requirement  & \multicolumn{2}{c}{14 TeV} & \multicolumn{2}{c}{27 TeV} \\
   \hline
    & SR-A & SR-B & SR-A & SR-B  \\
    \cline{2-5}
    $N$(muons/R-hadrons)
    & $\geq 1$ & $\geq 1$ & $\geq 1$ & $\geq 1$  \\
    $Z$-veto
    & \checkmark  & \checkmark  & \checkmark  & \checkmark  \\
    $|\eta|$ (rad) $<$
     & 2.4 & 2.4 & 2.4 & 2.4 \\
    $E_{T}^{\text{miss}} \text{(GeV)}>$
    & 90 & 90 & 120 & 120  \\
    $\Delta R(\text{track},\text{jet}_1) \text{(rad)}>$
    & 0.4 & 0.4 & 0.6 & 0.6 \\
    $\beta_s >$
    & 0.6 & 0.6 & 0.6 & 0.6 \\
    $\beta_s <$
    & 0.9 & 0.9 & 0.9 & 0.9 \\
    $p_T(\mu,R_{\tilde{t}})\text{ (GeV)}>$
    & 500 & 600 & 600 & 1200  \\
    \hline
     \end{tabulary}\end{center}
	\caption{The final cuts and preselection criteria used for the analysis of long-lived $R$-hadrons for each sub-signal regions SR-A and SR-B at 14 TeV and 27 TeV. }
	\label{tab4}
\end{table}

The SR is split into two sub-regions SR-A and SR-B corresponding to a variation in the cut imposed on the muon/$R$-hadron transverse momentum, $p_T(\mu,R_{\tilde{t}})$. The cut on this variable is optimized for the 14 TeV and 27 TeV studies as shown. It is natural to consider harder cuts on $p_T$ when looking at 27 TeV.
In Fig.~\ref{fig1} we exhibit the distributions in the variable $\beta_s$ for the benchmarks of Table~\ref{tab1} at 14 TeV (left panel) and 27 TeV (right panel) for 3000$\ifb$ of integrated luminosity.  One can clearly see that $\beta_s$ is peaked closer to one for lighter stops while it shifts for smaller values for heavier stops. Also it is evident that a cut on $\beta_s$ greater than 0.6 will remove a large part of the signal.

\begin{figure}[H]
 \centering
   \includegraphics[width=0.49\textwidth]{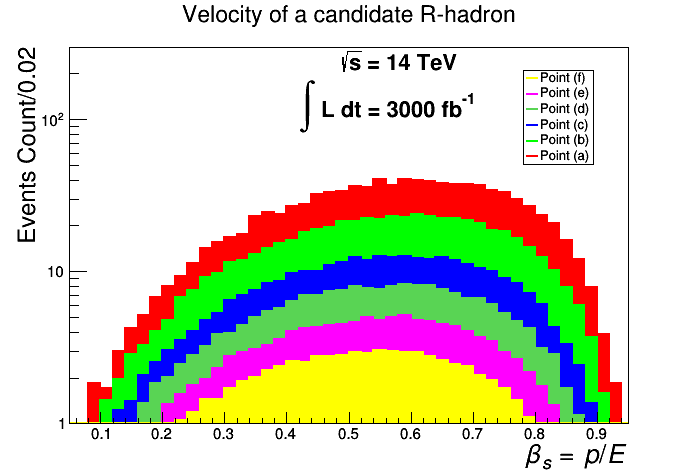}
      \includegraphics[width=0.49\textwidth]{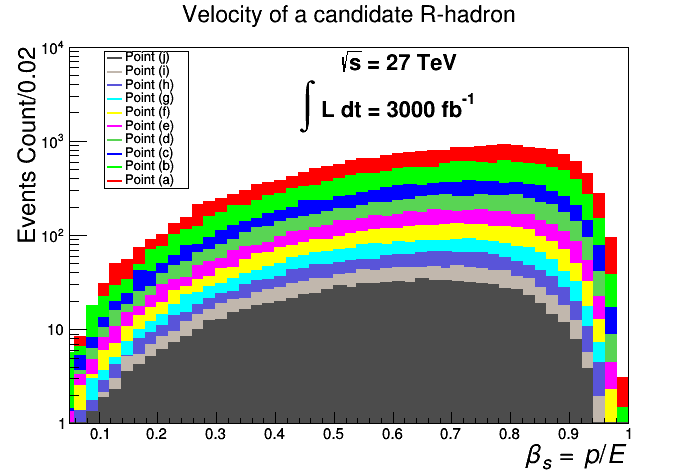}
   \caption{Distributions in the velocity $\beta_s$ of candidate $R$-hadrons at 14 TeV for points (a)$-$(f) (left panel) and 27 TeV for all points of Table~\ref{tab1} (right panel) both scaled to an integrated luminosity of 3000$\ifb$.}
	\label{fig1}
\end{figure}

After applying all the cuts in Table~\ref{tab4} except the cut on the transverse momentum of the muon/$R$-hadron we plot the distributions in this $p_T$ in Figs.~\ref{fig2} and~\ref{fig3} for the signal $S$ (black histogram) and the SM background $B$ (colored histograms). Actually we show $S$ versus $\sqrt{S+B}$ so that one can visually see the excess of the signal over the background. Thus Fig.~\ref{fig3} exhibits two signal points (a) and (c) of Table~\ref{tab1} which can be discovered at HL-LHC and HE-LHC, respectively. In the left panel, the signal and backgrounds are scaled to 300$\ifb$ at 14 TeV and one can see that a cut on $p_T$ greater than 500 GeV will remove most of the background. For point (c) in the right panel, a cut greater than $\sim$ 600$-$700 GeV is required for a $5\sigma$ discovery at a lesser integrated luminosity of 100$\ifb$ at 27 TeV. This shows the reason behind choosing those particular cuts in $p_T(\mu,R_{\tilde{t}})$ as shown in Table~\ref{tab4}.

\begin{figure}[H]
 \centering
   \includegraphics[width=0.49\textwidth]{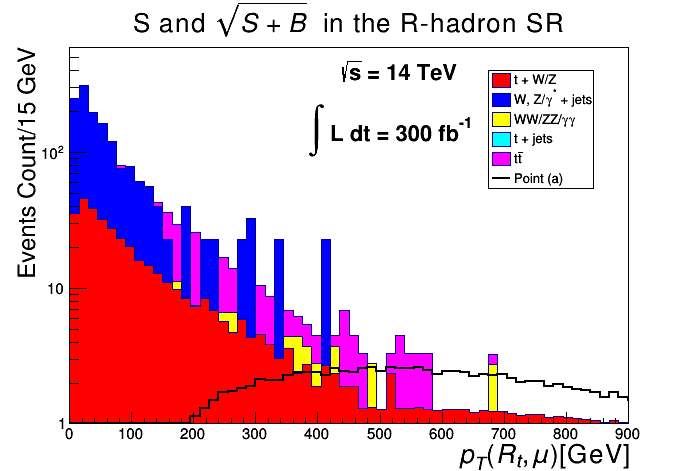}
   \includegraphics[width=0.49\textwidth]{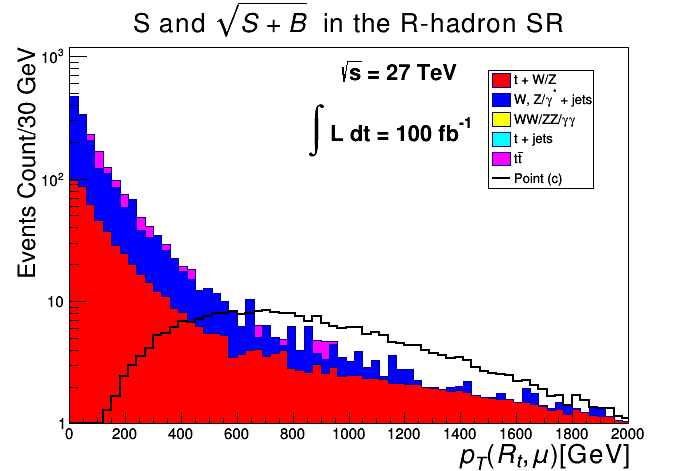}
   \caption{Left panel: Distributions in the transverse momentum of a stop $R$-hadron, $R_{\tilde{t}}$, of point (a) and of muons (SM backgrounds) at 14 TeV and 300$\ifb$ of integrated luminosity. Right panel: same as left panel but for point (c) at 27 TeV and 100$\ifb$.}
	\label{fig2}
\end{figure}

Benchmark points with larger stop masses have less chance of being discovered at HL-LHC. We show one such point in Fig.~\ref{fig3}, namely, point (e). Both panels show distributions in $p_T(\mu,R_{\tilde{t}})$ for the signal and backgrounds scaled to 500$\ifb$ but one at 14 TeV (left panel) and the other at 27 TeV (right panel). The signal is below the background for the entire $p_T$ range at 14 TeV while an excess can be seen beyond $\sim 600$ GeV at 27 TeV.

\begin{figure}[H]
 \centering
   \includegraphics[width=0.49\textwidth]{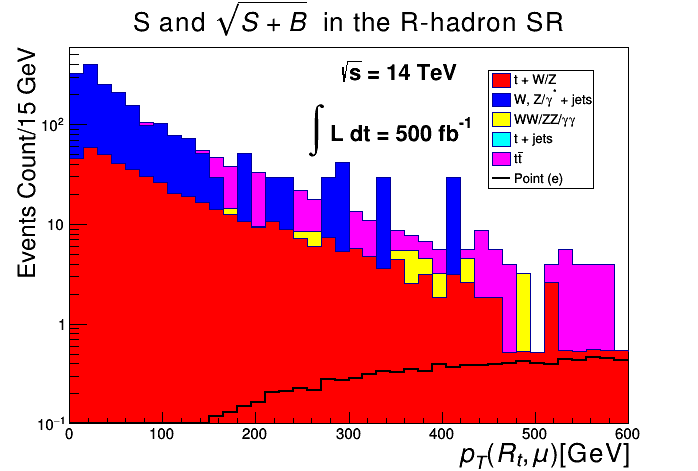}
   \includegraphics[width=0.49\textwidth]{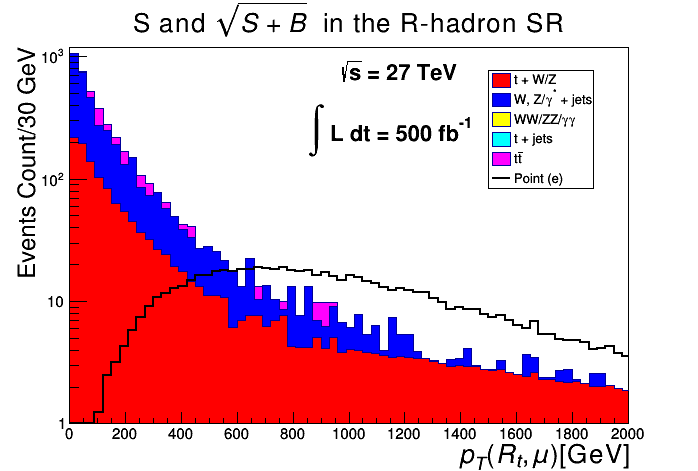}
   \caption{Comparison between distributions in $p_T$ for $R$-hadrons/muons at 14 TeV and 27 TeV for benchmark (e) at 500$\ifb$ of integrated luminosity.}
	\label{fig3}
\end{figure}

Applying all the cuts in Table~\ref{tab4} for our ten signal points (benchmarks of Table~\ref{tab1}) and SM backgrounds at 14 TeV and 27 TeV, we calculate the minimum integrated luminosity for $\frac{S}{\sqrt{S+B}}$ at the 5$\sigma$ level discovery. The results are shown in Table~\ref{tab5}.

\begin{table}[H]
	\centering
	\begin{tabulary}{\linewidth}{l|cc|cc}
    \hline\hline
	 & \multicolumn{2}{c}{$\mathcal{L}$ at 14 TeV} & \multicolumn{2}{c}{$\mathcal{L}$ at 27 TeV} \\
	\cline{2-5}
	Model & SR-A & SR-B & SR-A & SR-B \\
	\hline
  (a) & 259 & 226 & 20 & 21 \\
  (b) & 527 & 396 & 37 & 27 \\
  (c) & 1309 & 756 & 85 & 41 \\
  (d) & 2767 & 1226 & 150 & 55 \\
  (e) & $\cdots$ & 2128 & 308 & 81 \\
  (f) & $\cdots$ & 3667 & 591 & 119 \\
  (g) & $\cdots$ & $\cdots$ & 1258 & 189 \\
  (h) & $\cdots$ & $\cdots$ & 2387 & 285 \\
  (i) & $\cdots$ & $\cdots$ & 4831 & 461 \\
  (j) & $\cdots$ & $\cdots$ & 9922 & 791 \\
	\hline
	\end{tabulary}
	\caption{Comparison between the estimated integrated luminosity ($\mathcal{L}$) for a 5$\sigma$ discovery at 14 TeV (middle column) and 27 TeV (right column) for a stop $R$-hadron following the selection cuts, where the minimum integrated luminosity needed for a $5\sigma$ discovery is given in fb$^{-1}$. Entries with ellipses mean that the evaluated $\mathcal{L}$ is much greater than $3000\ifb$.}
\label{tab5}
\end{table}

The smallest integrated luminosities are obtained in the signal region SR-B which uses harder $p_T$ cuts. This is natural since $R$-hadrons are characterized by their large transverse momenta. Further, harder cuts on $p_T$ seem to produce better results especially for points with larger stop mass (points (g)$-$(j) at 27 TeV). For HL-LHC, a 1.4 TeV long-lived stop (point (a)) may be discoverable with an integrated luminosity as small as $\sim 230\ifb$, while point (e) will require $\sim 2000\ifb$. Points (f)$-$(j) appear to be out of reach of HL-LHC as they require more than 3000$\ifb$. At the HE-LHC, the entire stop mass range (1.4 TeV to 2.3 TeV) appears to be within reach requiring an integrated luminosity as low as 20$\ifb$ for point (a) and $\sim 800\ifb$ for point (j) for discovery. For a visual comparison, the results from SR-B are displayed in Fig.~\ref{fig4} with the left panel showing the points that are discoverable at both HL-LHC and HE-LHC while the right panel shows the rest of the points which are only discoverable at HE-LHC.

\begin{figure}[H]
 \centering
   \includegraphics[width=0.49\textwidth]{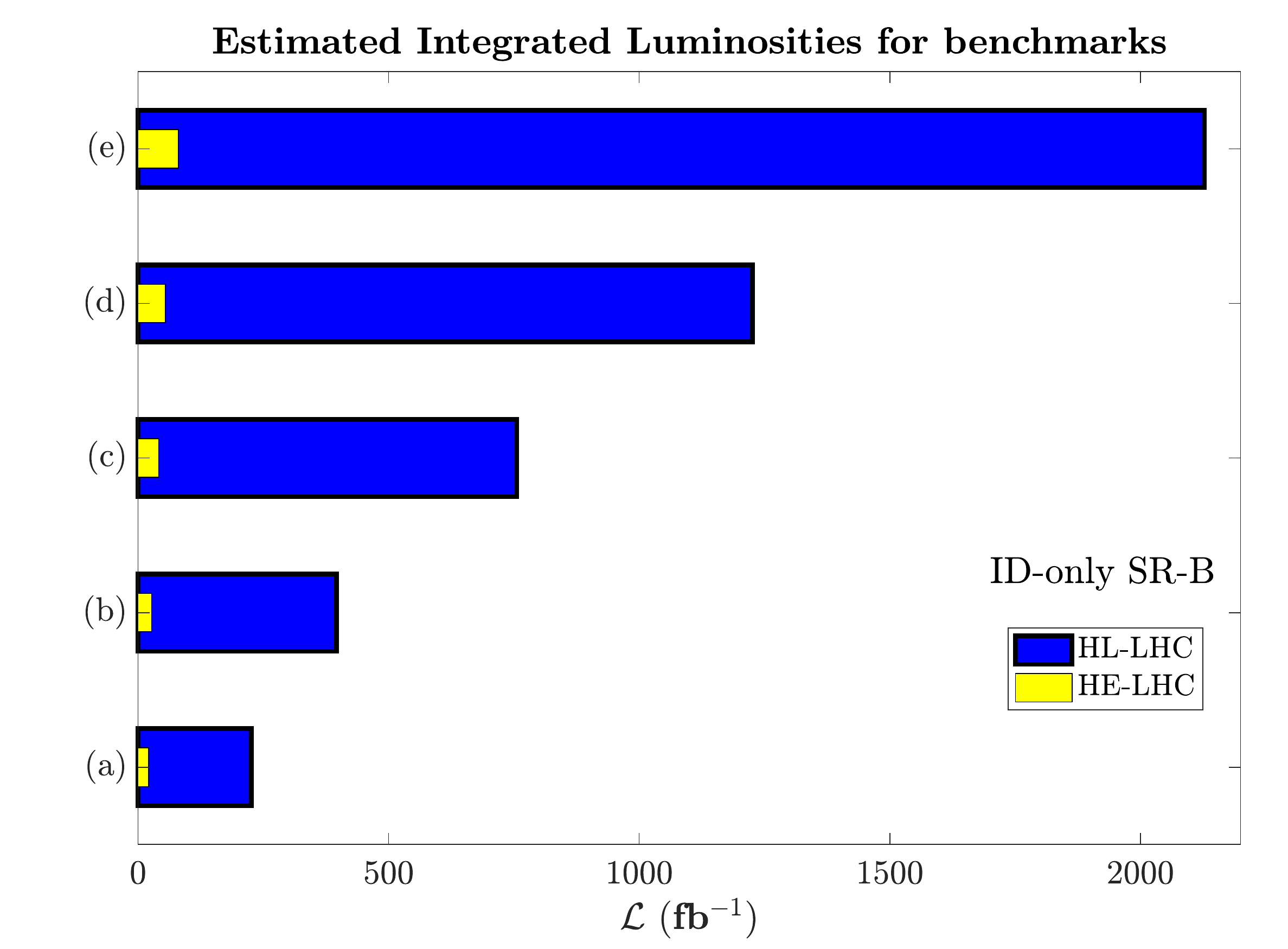}
   \includegraphics[width=0.49\textwidth]{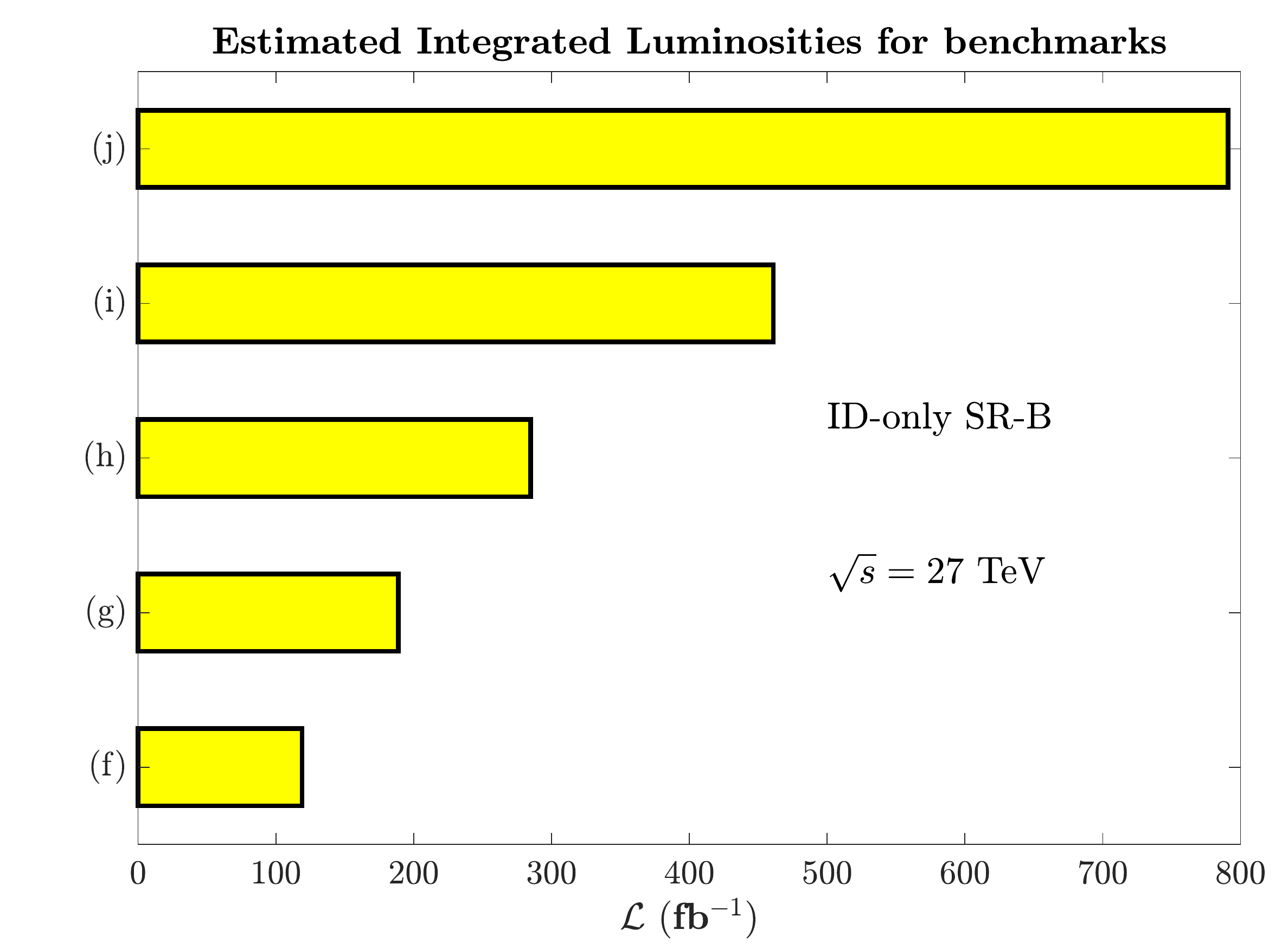}
   \caption{Left panel: the integrated luminosity for discovery of the points (a)$-$(e) which are discoverable at both HL-LHC and HE-LHC. Right panel: the integrated luminosity for discovery of the points (f)$-$(j) at HE-LHC. }
	\label{fig4}
\end{figure}

As a comparison between HL-LHC and HE-LHC, we estimate a time frame for discovery using the rates at which HL-LHC and HE-LHC will be collecting data. For the HL-LHC, point (a) may be discoverable within $\sim 8$ months from resuming operation while points (b)$-$(e) will require a period of $\sim 1.2$ yrs to $\sim 7$ yrs. For HE-LHC, it is expected that such a machine will collect data at a rate of 820$\ifb$/yr and so points (a)$-$(d) will require $\sim 9$ to 24 days of runtime while the rest of the points will take $\sim 1$ yr to 12 yrs of runtime for a potential discovery. The advantage of switching to a 27 TeV collider is evident in terms of its mass reach capabilities as well as reducing the runtime for discovery of SUSY.

We clarify further the connection of cosmology and collider phenomenology discussed above.
The analysis of this work is based on the assumption that the stop is long-lived and leaves a track inside the detector as an $R$-hadron which acts like a heavy muon and then decays outside the detector into the hidden sector neutralino and contributes to its relic density. This is what connects cosmology to the collider phenomenology.
Since the stop is long-lived and decays outside the detector,
a further test of this model could come about by detection of its decay in future detectors
which would have the ability at exploring the lifetime frontier.
MATHUSLA~\cite{Lubatti:2019vkf} and FASER~\cite{Ariga:2019ufm} are examples of such detectors
capable of detecting long-lived particles which decay further away from their production vertex.
Thus a detection of the stop track inside ATLAS or CMS
along with future detectors far enough to detect the decay products given the long lifetime of stop
would lend support to the underlying model proposed here which connects cosmology to collider physics.

\section{Comments and caveats on the connection between cosmology and LHC phenomenology}
Here we discuss the caveats that relate dark matter with the LHC phenomenology of the model in this work.
First we discuss the possibility that stop may be the LSP of the whole system but that it  annihilates rapidly so it
no longer contributes any discernible  amount to the relic density  of dark matter in the Universe.
In this circumstance dark matter
would be  disconnected from the particle physics phenomenology  at the LHC.
 We examined this possibility in the context of the current experimental  limits on the heavy charged particles $X^+$.
 The limits on the yield
 of such heavy charged particles in deep sea water experiment  (including  gravitational effects)
 with  masses in the range
  5 GeV $\leq m_{X^+} \leq$ 1.6 TeV
 is given by~\cite{Yamagata:1993jq} (see also the related works~\cite{Smith:1982qu,Hemmick:1989ns,Verkerk:1991jf,Norman:1988fd})
 \begin{align}
 Y_{X^+}\leq 0.9 \times 10^{-38}
  \left(\frac{\Omega_B h^2} {0.0223}\right),
  \label{9.1}
  \end{align}
  which corresponds to a concentration of the order $10^{-28}$ at the sea level.  For larger masses in the range
  10 TeV $\leq m_{X^+} \leq 6\times 10^4$  TeV  the limits are
  \begin{align}
   Y_{X^+}\leq 6 \times 10^{-25}\left(\frac{\Omega_B h^2} {0.0223}\right).
   \label{9.2}
   \end{align}
  Such small yields cannot be obtained in any
  reasonable manner in MSSM even if we saturate the unitarity bound on the annihilation cross section.
  To illustrate this point more concretely, we have carried out a
  scan of the parameter space of the MSSM looking for points where the stop is the LSP and using the Higgs boson mass constraint. A scatter plot is shown in Fig.~\ref{fig5}. We find that the stop yield is a factor of order $\sim 10^{10}$ or more  larger than the current experimental bound of Eq.~(\ref{9.2}).
 A similar conclusion is reached in the work of~\cite{Berger:2008ti}
 which states that the experimental bounds on heavy charged relics
 are so strong that the possibility of such a relic to be dark matter is completely excluded.   Our analysis shows that at least
 for the case of MSSM/SUGRA model, the stop being an LSP consistent with the current
 experimental limits of deep sea water is not feasible.

\begin{figure}[H]
 \centering
   \includegraphics[width=0.8\textwidth]{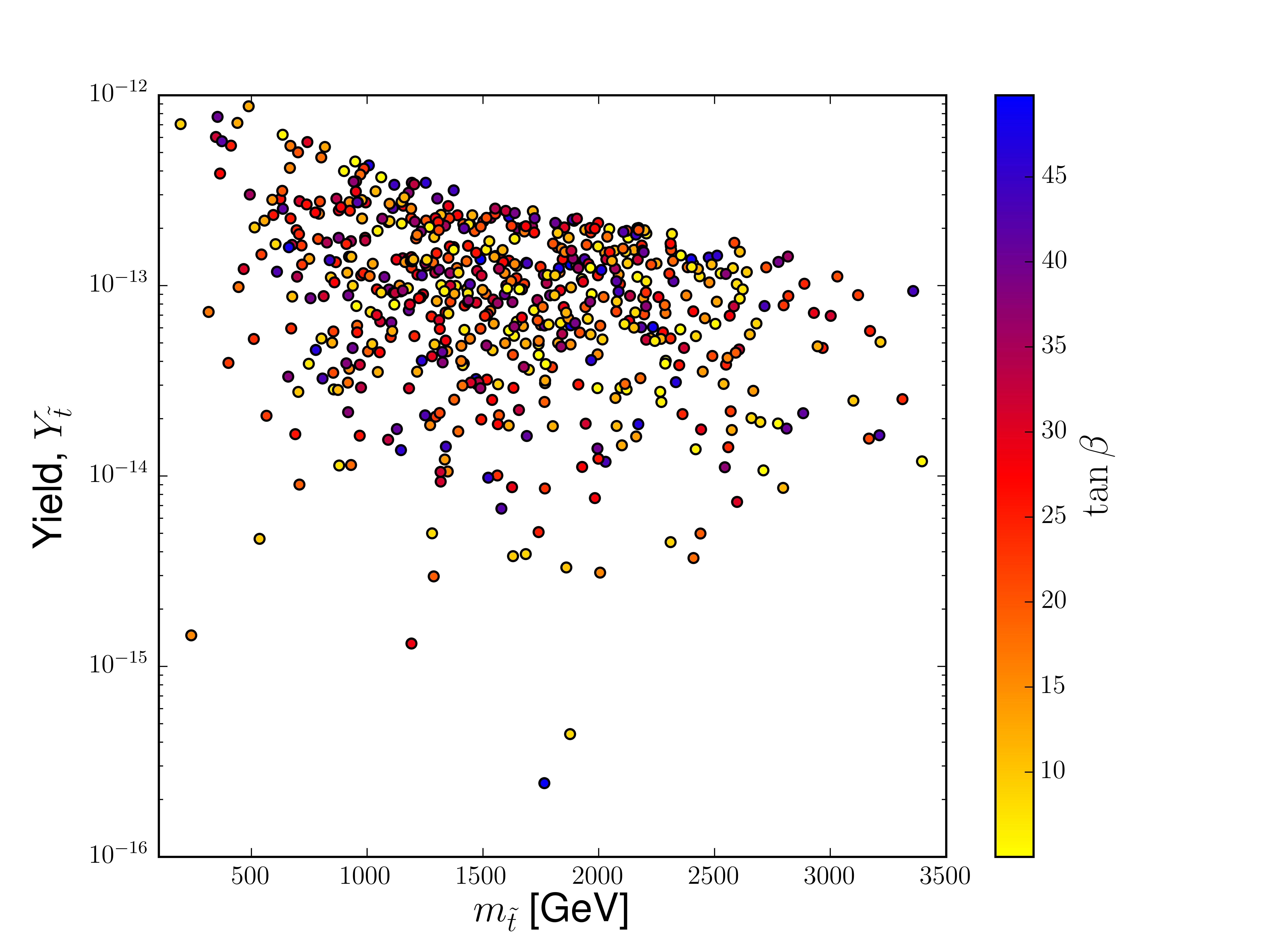}
   \caption{A scatter plot of the stop yield versus the stop mass with $\tan\beta$ shown on the color axis. The range of the SUGRA parameters used in the scan: $m_0\in [300,8000]$, $A_0/m_0\in [-4,4]$, $m_1, m_2\in [2000,8000]$, $m_3\in [1000,8000]$ and $\tan\beta\in [5,50]$ with sgn$(\mu)>0$.}
	\label{fig5}
\end{figure}

We note here  that a firm test of the proposed model would be the detection of decay of a long-lived stop
in  future particle detectors such as MATHUSLA.  Such an analysis would involve simulations of long-lived particle
detectors not yet built and  is outside the framework of the current work but is an interesting topic for a  future project.
Finally we discuss various caveats  connecting dark matter
and LHC phenomenology. Such a connection is highly model dependent. For instance if the relic density of the LSP of the visible sector
could be depleted to be consistent with the current limits on massive charged particles as given by  experiment on deep sea water,
the dark matter particle could be something else such as an axion or some other hidden sector particle and there would be no relation
between the existence of dark matter and the particle phenomenology at the LHC.

In summary our analysis is a very specific one
based on MSSM/SUGRA model where the couplings are highly constrained by supersymmetry.  Thus for example, the
annihilation of the stops in our model takes place dominantly via  Higgs boson $h$, $Z$ and $Z'$ direct channel poles. Their couplings
are constrained by gauge invariance and by supersymmetry.  Consequently the allowed values of the annihilation cross sections are
 constrained. Additionally the SUSY parameters are constrained by current lower limits on sparticle masses and by direct and
 indirect detection experiments. Within these constraints the stop  being the LSP of the entire model is not feasible.
  Thus a robust prediction of the model is a long-lived stop which would decay outside the detector. The possibility of testing
  this model exists in future long-lived particle detectors.

\section{Conclusions}\label{sec:conc}

In this work we  discussed the possibility that the neutralino  in the hidden sector is the lightest supersymmetric particle,
  and specifically lighter than all the sparticles in the MSSM spectrum. Further we assume that the hidden sector neutralino
  interacts with the visible sector with ultraweak interactions. In this case all the  sparticles in MSSM  will eventually decay to the hidden sector neutralino which
  will be a dark matter candidate. We investigate this possibility in a concrete setting. We consider an $U(1)_X$ gauge extension of MSSM/SUGRA model which
  will have two $U(1)$ gauge factors:  $U(1)_X$ and $U(1)_Y$ where $U(1)_Y$ is the gauge group of the hypercharge.  Here one has
  the possibility of gauge kinetic mixing and Stueckelberg mass mixing between the two $U(1)$
  gauge groups. If the mixing between the two is very small, one
  has interactions between the hidden sector and the visible sector which are ultraweak.
  In this case the LSP in the MSSM sector  will decay into the hidden sector neutralino with a long lifetime  and
   will escape the detector without decay and if charged it will
    leave a track inside the detector. In the analysis below we investigate concrete models where this situation is realized.
  Specifically we consider models where the LSP in the MSSM sector is a stop which decays into the hidden sector dominantly via the process $\tilde t \to \tilde\xi^0_1~t$, where $\tilde\xi^0_1$ is the dark matter particle in the hidden sector.

In the analysis presented here we investigate a set of benchmarks containing a stop NLSP  with mass range of 1.4 TeV to 2.3 TeV which is long-lived and
  carry out a collider analysis for its  discovery at the HL-LHC and HE-LHC.  A long-lived stop hadronizes  into an $R$-hadron
   made up of the stop parton surrounded by light standard model quarks. The $R$-hadron is color neutral but electrically charged and
   can be identified by the track it leaves in the detector.  It  is characterized by its large transverse momentum and slow speed $\beta_s$.
   In our analysis we focused on information from the tracker and we showed that half of the benchmarks of Table \ref{tab1}
   corresponding to a stop in the mass range 1.4 TeV to 1.8 TeV can be discovered at HL-LHC while all the benchmarks of Table \ref{tab1}
    are discoverable at HE-LHC. At HL-LHC, an integrated luminosity $\sim 230\ifb$ is needed to discover a 1.4 TeV stop which is right around the
    corner once the LHC is back to collecting more data. The integrated luminosity for discovery is greatly reduced at HE-LHC where
   an integrated luminosity  as low as 20$\ifb$ is sufficient to discover a 1.4 TeV stop and an integrated luminosity of
      $\sim 800\ifb$ is sufficient to discover a 2.3 TeV stop.

An important conclusion of our analysis is that even for dark matter with ultraweak or feeble interactions, the freeze-in relic density is not an accurate
measure of the total relic density and
one must include the freeze-out contribution from the next-to-lightest supersymmetric particle.
Thus our analysis based on the benchmarks of Table~\ref{tab1} and Table~\ref{tab2} shows
that freeze-in relic density is typically dominant for part of the parameter space where the stop masses are relatively small where its relative contribution to the total
relic density can be up to $\sim 70\%$ (for model (c)) but
is typically subdominant for relatively large stop masses where its contribution is as small as  only $\sim 24\%$ (for model (j)).

\vspace{1cm}

\textbf{Acknowledgments: }
The analysis presented here was done using the resources of the high-performance Cluster353 at the Advanced Scientific Computing Initiative (ASCI) and the Discovery Cluster at Northeastern University. WZF is grateful to Wei Chao, Xiaoyong Chu for helpful discussions and WZF also thanks Northeastern University for hospitality. WZF was supported in part by the National Natural Science Foundation of China, the Youth Science Fund under Grant No. 11905158 and Tianjin construction of high-end talents Fund.
The research of AA and PN was supported in part by the NSF Grant PHY-1913328.

\end{document}